\documentclass[twocolumn,aps,pra,floatfix,superscriptaddress,notitlepage]{revtex4-1}
\usepackage[utf8]{inputenc}
\usepackage[OT1]{fontenc}
\usepackage{amsmath}
\usepackage{amsfonts}
\usepackage{amssymb}
\usepackage{bm}
\usepackage{graphicx}
\usepackage[left=2cm,right=2cm,top=2cm,bottom=2cm]{geometry}
\usepackage{longtable,booktabs}
\allowdisplaybreaks
\usepackage{dcolumn}
\usepackage{siunitx}
\usepackage{multirow}
\usepackage{bigdelim}

\usepackage{color}
\definecolor{BLUE}{rgb}{0.0,0.0,1.0}

\usepackage{soul}
\soulregister\cite7
\soulregister\ref7

\newcommand{\be}{\begin{eqnarray}}
\newcommand{\ee}{\end{eqnarray}}

\begin{document}

\title{QED calculations of intra-$L$-shell singly excited states in Be-like ions}

\author{A.~V.~Malyshev}
\affiliation{Department of Physics, St.~Petersburg State University, Universitetskaya 7-9, 199034 St.~Petersburg, Russia  
\looseness=-1}
\affiliation{Petersburg Nuclear Physics Institute named by B.P. Konstantinov of National Research Center ``Kurchatov Institute'', Orlova roscha 1, 188300 Gatchina, Leningrad region, Russia}

\author{Y.~S.~Kozhedub}
\affiliation{Department of Physics, St.~Petersburg State University, Universitetskaya 7-9, 199034 St.~Petersburg, Russia  
\looseness=-1}

\author{V.~M.~Shabaev}
\affiliation{Department of Physics, St.~Petersburg State University, Universitetskaya 7-9, 199034 St.~Petersburg, Russia  
\looseness=-1}
\affiliation{Petersburg Nuclear Physics Institute named by B.P. Konstantinov of National Research Center ``Kurchatov Institute'', Orlova roscha 1, 188300 Gatchina, Leningrad region, Russia}

\author{I.~I.~Tupitsyn}
\affiliation{Department of Physics, St.~Petersburg State University, Universitetskaya 7-9, 199034 St.~Petersburg, Russia  
\looseness=-1}


\begin{abstract}

The \textit{ab initio} approach is used to evaluate the excitation energies of the $2s2p \, ^{2S+1}P_J$ states from the ground state as well as the $2s2p \, ^3P_1 \rightarrow 2s2p \, ^3P_0$ and $2s2p \, ^3P_2 \rightarrow 2s2p \, ^3P_1$ transition energies for selected Be-like highly-charged ions over a wide range: from ${\rm Ar}^{14+}$ to ${\rm U}^{88+}$. The issue of a strong level mixing due to the proximity of states with the same symmetry is addressed by applying the QED perturbation theory for quasidegenerate levels. The employed approach combines a rigorous perturbative QED treatment up to the second order with electron-electron correlation contributions of the third and higher orders calculated in the Breit approximation. The higher-order QED effects are estimated using the model-QED-operator approach. The nuclear-recoil and nuclear-polarization effects are taken into account as well. The performed calculations are accompanied with a thorough analysis of uncertainties due to uncalculated effects. The most accurate theoretical predictions for the excitation and transition energies in Be-like ions are obtained, which, in general, are in perfect agreement with the available experimental data. 

\end{abstract}


\maketitle


\section{Introduction \label{sec:0}}

Quantum electrodynamics (QED) is widely recognized as a powerful tool for describing the electronic structure and various properties of bound systems composed of electrically charged particles. The strengths of this theory are clearly manifested in the study of highly-charged ions (HCIs)~\cite{Beyer:2003:book:eng}. In HCIs, non-trivial QED effects are significantly enhanced compared to those in light atoms. On the other hand, these effects are not masked by an uncertainty with which electron-electron correlation effects can be treated, as is usually the case for neutral systems. All this makes HCIs ideal candidates for a diverse range of fundamental investigations. For example, high-precision measurements with HCIs traditionally form the basis for comprehensive tests of bound-state-QED methods~\cite{Sapirstein:2008:25, Beiersdorfer:2010:074032, Glazov:2011:71, Volotka:2012:073001, Shabaev:2015:031205, Shabaev:2018:60, Kozlov:2018:045005, Indelicato:2019:232001, Shabaev:2024:94:inbook}. Furthermore, HCIs can also be used for the precise determination of fundamental constants~\cite{Sturm:2014:467, Shabaev:2006:253002, Volotka:2014:023002, Yerokhin:2016:100801}, in the search for their possible variation \cite{Andreev:2005:243002, Berengut:2010:120801, Oreshkina:2017:030501}, and in numerous other applications~\cite{Shabaev:2022:043001, King:2022:43}.

To date, one of the most rigorous tests for bound-state-QED methods is associated with the measurement of the $2p_{1/2} \rightarrow 2s$ transition in Li-like uranium~\cite{Schweppe:1991:1434, Brandau:2003:073202, Beiersdorfer:2005:233003}, for related theory see, e.g., Refs.~\cite{Kozhedub:2008:032501, Sapirstein:2011:012504}. In our recent work~\cite{Malyshev:2021:183001}, which was primarily devoted to QED calculations for Xe$^{50+}$, we also evaluated the $2p_{1/2} \rightarrow 2s$ transition in Be-like uranium. The theoretical accuracy achieved was found to be comparable to that for Li-like uranium. As the measurements of these systems can also be performed at a comparable level of accuracy~\cite{Beiersdorfer:2005:233003}, both these charge states become equally attractive for probing bound-state QED in the strong-coupling regime. A similar conclusion was reached in Ref.~\cite{Malyshev:2023:042806}, where the $2p_{3/2} \rightarrow 2s$ transition energies in He-, Li-, and Be-like uranium, as well as all their pairwise differences, were calculated. The results reported in Ref.~\cite{Malyshev:2023:042806} demonstrate good agreement with the high-precision measurements, accomplished recently at the ESR storage ring at the GSI in Darmstadt~\cite{Loetzsch:2024:673}.

At the same time, bound-state-QED calculations of Be-like ions are in certain respects more challenging than those of Li-like ions. In Be-like ions, the proximity of energy levels with the same symmetry results in their strong mixing by the electron-electron interaction. As a consequence, the conventional single-level QED perturbative approach, which appears to be valid for Li-like ions~\cite{Kozhedub:2008:032501, Sapirstein:2011:012504}, may prove inadequate when applied to Be-like ions~\cite{Malyshev:2021:183001}. This strong mixing of energy levels inevitably arises, e.g., in the study of the $2s2s \, ^1S_0$ ground state of Be-like ions~\cite{Armstrong:1976:1114, Braun:1984:book:rus2eng} (for the sake of brevity, we do not include the closed $K$ shell into the state designations). The intra-$L$-shell singly excited states with the total angular momentum $J=1$ are also significantly mixed by the electron-electron interaction. 

The investigation of the electronic structure of Be-like HCIs has been addressed many times previously by means of a variety of relativistic approaches~\cite{Armstrong:1976:1114, Cheng:1979:111, Edlen:1983:51, Lindroth:1992:2771, Marques:1993:929, Zhu:1994:3818, Safronova:1996:4036, Chen:1997:166, Santos:1998:149, Cheng:2000:054501, Safronova:2000:1213, Majumder:2000:042508, Dong:2001:294, Draganic:2003:183001, Gu:2005:267, SoriaOrts:2006:103002, Ho:2006:022510, Cheng:2008:052504, Winters:2011:014013, Sampaio:2013:014015, Yerokhin:2014:022509, Yerokhin:2015:054003, Wang:2015:16, El-Maaref:2015:2, Li:2017:720, Kaygorodov:2019:032505}. However, to the best of our knowledge, all these calculations treated the QED effects at best within some one-electron (first-order) approximations only, demonstrating a significant scatter of the resulting theoretical predictions. In contrast, the method developed by us in Ref.~\cite{Malyshev:2021:183001} takes into account all the relevant QED contributions up to the second order, overcoming the challenges due to the level mixing by using perturbation theory (PT) for quaidegenerate levels. The first implementation of this method for Be-like HCIs was realized in Refs.~\cite{Malyshev:2021:183001, Malyshev:2021:652, Malyshev:2023:042806} for the ground and low-lying excited states in xenon ion, as well as for selected transitions in molybdenum and uranium ions. The present work extends these calculations to a broader range of Be-like HCIs: from argon to uranium. Namely, we evaluate the excitation energies of the $2s2p \, ^3P_{0,1,2}$ and $2s2p \, ^1P_1$ states from the $2s2s \, ^1S_0$ ground state. The $2s2p \, ^3P_1 \rightarrow 2s2p \, ^3P_0$ and $2s2p \, ^3P_2 \rightarrow 2s2p \, ^3P_1$ transition energies are also considered. In all of the cases, particular attention is paid to a comprehensive examination of the uncertainties associated with uncalculated effects. The obtained theoretical predictions are compared with the available experimental data~\cite{Widing:1975:L33, Dere:1978:1062, Denne:1989:1488, Denne:1989:3702, Martin:1990:6570, Sugar:1991:859, Beiersdorfer:1993:3939, Beiersdorfer:1995:2693, Bieber:1997:64, Beiersdorfer:1998:1944, Trabert:2003:042501, Feili:2005:48, Katai:2007:120, Saloman:2010:033101, Bernhardt:2015:144008} and the results of the previous calculations. In general, good agreement is found with the former.



\section{Theoretical approach and computational details \label{sec:1}}

In this section, we briefly describe the \textit{ab initio} method employed in the present work to evaluate the electronic structure of Be-like HCIs. Some numerical aspects of the calculations are discussed here as well. For further details, the reader is referred to Refs.~\cite{Malyshev:2021:183001, Malyshev:2021:652, Malyshev:2023:042806} and references therein.

We use PT which is constructed in the Furry picture of QED~\cite{Furry:1951:115} according to the two-time Green’s function (TTGF) method~\cite{TTGF}. To zeroth order, this means that individual electrons obey the Dirac equation with some local spherically symmetric binding potential~$V$. To better keep under control the accuracy of the calculations and to analyze the convergence of the perturbation series, we perform all calculations for several choices of~$V$. First, we take $V$ to be the Coulomb potential of an extended nucleus, $V=V_{\rm nucl}$. Second, we modify~$V$ by adding some screening potential to the nuclear one, $V=V_{\rm nucl}+V_{\rm scr}$. The latter is introduced to partially take into account the interelectronic-interaction effects from the very beginning. Obviously, this modification leads to a rearrangement of perturbation series, since the counterterm $\delta V = -V_{\rm scr}$ is to be treated perturbatively. As in Ref.~\cite{Malyshev:2021:183001}, we use two variants for the screening potential: (i) the core-Hartree (CH) potential generated by the closed $K$ shell; and (ii) the local Dirac-Fock (LDF) potential~\cite{Shabaev:2005:062105}. 

As noted in Sec.~\ref{sec:0}, for a proper description of the electronic structure of Be-like HCIs, it is crucial to take into account the proximity of energy levels possessing the same symmetry. Within the TTGF method this issue is solved by applying the QED PT for quasidegenerate levels~\cite{TTGF}. The close-lying levels are grouped into a finite-dimensional model subspace~$\Omega$, and an effective Hamiltonian~$H_{\rm eff}$, acting in this subspace, is constructed step-by-step including all relevant contributions. The energies of the states under consideration are obtained by diagonalizing the matrix of~$H_{\rm eff}$. In this respect, the standard QED PT for a single level can be considered as a special case with the one-dimensional~$\Omega$. It is important to note a peculiarity of PT with the dimension of the model subspace greater than one: the contributions due to different effects are not strictly additive, as they are mixed during the diagonalization. Therefore, this diagonalization must be performed at the last stage of the calculations.

Before proceeding with the discussion of the model subspaces used in the present work, let us comment on the notations. Our \textit{ab initio} method represents the fully relativistic approach. All the unperturbed many-electron wave functions are constructed in the $jj$ coupling from Slater determinants, which are in turn built from the solutions of the one-electron Dirac equation. However, to emphasize the impact of energy-level mixing, the $LS$-coupling notation is utilized to refer to the states resulting from the $H_{\rm eff}$-matrix diagonalization procedure. For the sake of uniformity, we apply this also to the states obtained using one-dimensional model subspaces. More specifically, the $(2s2p_{1/2})_0$ and $(2s2p_{3/2})_2$ levels turn out to be separated from other levels of the same symmetry. Therefore, their energies are calculated using PT for a single level. The corresponding states are designated with $2s2p \, ^3P_0$ and $2s2p \, ^3P_2$, respectively. The $(2s2p_{1/2})_1$ and $(2s2p_{3/2})_1$ levels exhibit strong mixing with each other. We employ the two-dimensional model subspace~$\Omega$ spanned by these levels. The resulting states are referred to as $2s2p \, ^3P_1$ and $2s2p \, ^1P_1$. The influence of level mixing on the QED effects can be identified by recalculating the energies of the $(2s2p_{1/2})_1$ and $(2s2p_{3/2})_1$ levels regarding them as the single ones. In the present paper, the energies of the intra-$L$-shell singly excited states are given relative to the $2s2s \, ^1S_0$ ground state. Therefore, an accurate description of the latter is a vital task too. As in Refs.~\cite{Malyshev:2021:183001, Malyshev:2021:652}, we use the three-dimensional model subspace spanned by the $(2s2s)_0$, $(2p_{1/2}2p_{1/2})_0$, and $(2p_{3/2}2p_{3/2})_0$ levels. In addition, we study the two-dimensional $\Omega$ spanned by only the first two of these levels and the one-dimensional $\Omega$ corresponding to the $(2s2s)_0$ level. By combining the results of these three distinct calculations for the ground state, we gain a deeper understanding of the strong interplay between the electron-electron correlation and QED effects, which was noted in Ref.~\cite{Malyshev:2021:183001}. 

\begin{figure}
\begin{center}
\includegraphics[width=\columnwidth]{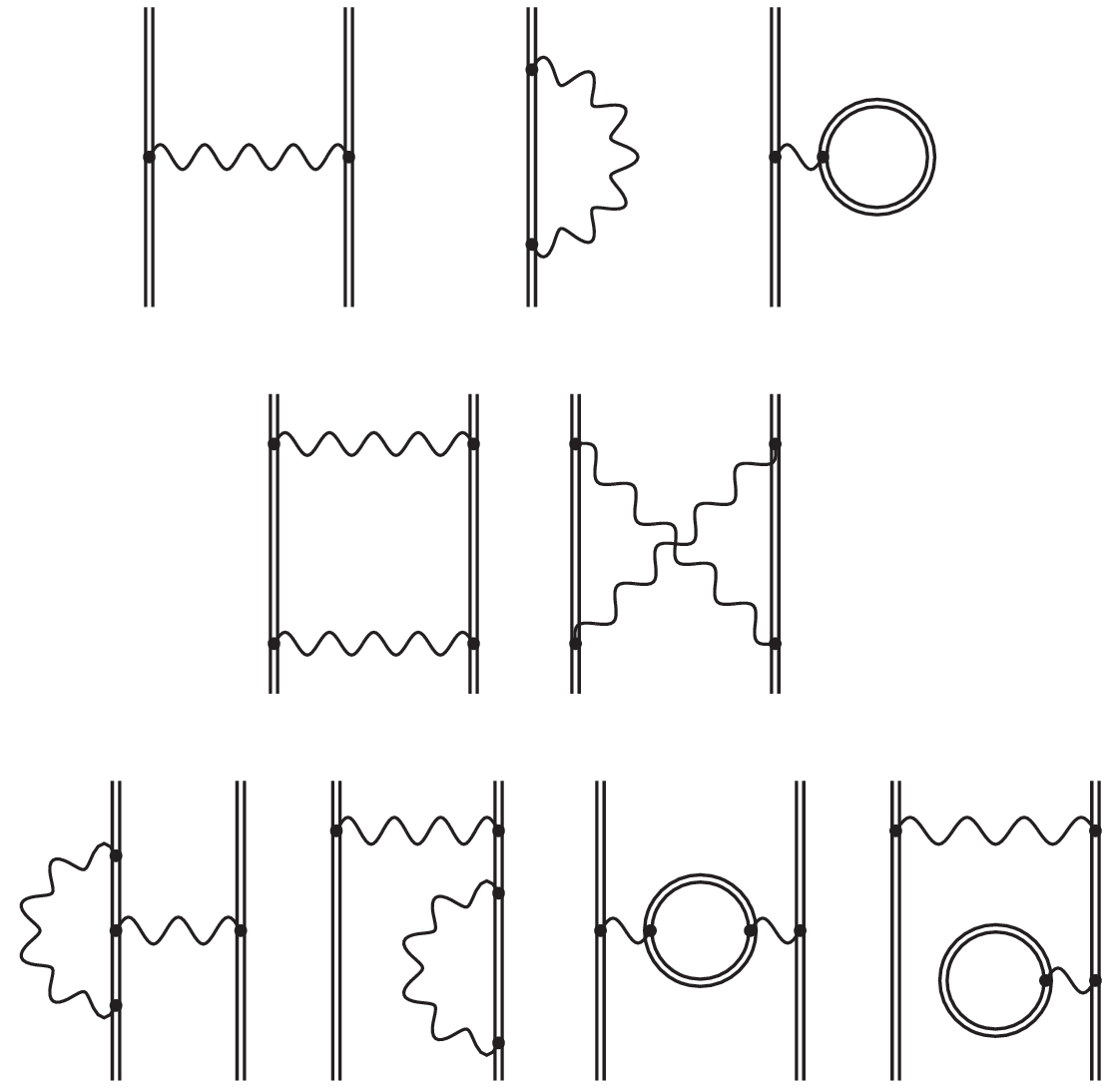}
\caption{\label{fig:diagr}
First- and second-order Feynman diagrams, except for the one-electron two-loop diagrams. The counterterm diagrams are not shown. The double line corresponds to the electron propagator in the binding potential~$V$. The formalism, in which the $1s$ state is assigned to the vacuum, is used~\cite{TTGF}. The wavy line represents the photon propagator.}
\end{center}
\end{figure}

The developed approach rigorously treats all contributions up to the second order of the QED PT. The computational expressions are derived within the TTGF method in the formalism, in which the closed $K$ shell is assigned to the vacuum~\cite{TTGF}. In this formalism, only two $L$-shell electrons remain, and, therefore, the problem is reduced to the consideration of one- and two-electron Feynman diagrams. The necessary diagrams, except for the one-electron two-loop diagrams and the counterterm diagrams, which arise when $V_{\rm scr}$ is included into the zeroth-order Hamiltonian, are shown in Fig.~\ref{fig:diagr}. The derivation of the formulas is performed for an arbitrary size of the model subspace~$\Omega$. By passing to the standard QED vacuum, we explicitly extract from the obtained formulas the contributions describing the interaction with the $K$-shell electrons. The calculations are organized as described in Ref.~\cite{Malyshev:2023:042806}: (i) the one-electron two-loop QED corrections are obtained using the results of Refs.~\cite{Yerokhin:2015:033103, Yerokhin:2018:052509}; and (ii) all other contributions are rigorously evaluated for the three binding potentials~$V$ listed above. Further details of the QED calculations can be found in our previous works, see, e.g., Refs.~\cite{Malyshev:2014:062517, Malyshev:2017:022512, Kozhedub:2019:062506, Malyshev:2019:010501_R, Malyshev:2021:183001, Malyshev:2021:652, Malyshev:2023:042806}.

The higher-order interelectronic-interaction effects are treated within the Breit approximation based on the Dirac-Coulomb-Breit (DCB) Hamiltonian. There are a variety of well-established methods for solving the DCB equation. In our approach, we use the configuration-interaction (CI) method in the basis of the Dirac-Sturm orbitals~\cite{Bratzev:1977:2655, Tupitsyn:2003:022511}. The key point is to accurately merge the results of QED and electron-electron correlation effects. While this issue is relatively straightforward in the case of PT for a single level~\cite{Kozhedub:2010:042513, Malyshev:2017:022512}, it becomes more complex when considering PT for quasidegenerate levels. For this aim, we utilize the procedure initially proposed in Ref.~\cite{Malyshev:2019:010501_R} and described in details in Refs.~\cite{Kozhedub:2019:062506, Malyshev:2021:652}.

To complete the brief description of the method, it should be noted that the corrections arising from the nuclear-recoil~\cite{Shabaev:1985:394, Shabaev:1988:107, Shabaev:1998:59, Pachucki:1995:1854, Adkins:2007:042508, Palmer:1987:5987, Pachucki:2024:032804} and nuclear-polarization~\cite{Plunien:1995:1119:1996:4614:join_pr, Nefiodov:1996:227, Yerokhin:2015:033103} effects are also taken into account. In addition, we employ the model-QED operator~\cite{Shabaev:2013:012513, Shabaev:2015:175:2018:69:join_pr} to estimate the higher-order screened QED contributions, which correspond to the Feynman diagrams similar to those in the last line in Fig.~\ref{fig:diagr} but with an additional photon line mediating the electron-electron interaction. The details of all these calculations are summarized in Ref.~\cite{Malyshev:2023:042806}.

The calculations are performed for the following Be-like HCIs: argon~$^{40}{\rm Ar}^{14+}$, krypton~$^{84}{\rm Kr}^{32+}$, molybdenum~$^{98}{\rm Mo}^{38+}$, xenon~$^{132}{\rm Xe}^{50+}$, lead~$^{208}{\rm Pb}^{78+}$, bismuth~$^{209}{\rm Bi}^{79+}$, thorium~$^{232}{\rm Th}^{86+}$, and uranium~$^{238}{\rm U}^{88+}$. The nuclear masses and root-mean-square radii are assumed to be the same as in Ref.~\cite{Yerokhin:2015:033103}. The nuclear-charge distribution is described by the Fermi model with the thickness parameter set to 2.3~fm. For uranium, we also consider the correction caused by the nuclear-deformation effect~\cite{Kozhedub:2008:032501}. The hyperfine structure due to the nuclear-magnetic moment of $^{209}{\rm Bi}$ is neglected. The values of the fundamental constants from Ref.~\cite{Tiesinga:2021:025010} are employed throughout the calculations. 


\section{Numerical results and discussions \label{sec:2}}

\begin{table*}[t]
\centering

\renewcommand{\arraystretch}{1.25}

\caption{\label{tab:amplitude} 
         Expansion coefficients, $A_i$, $B_i$, and $C_i$, of the many-electron wave functions in the configuration-state functions:
         $\Psi[2s2s\,^1S_0]=A_1\Phi[(2s2s)_0]+A_2\Phi[(2p_{1/2}2p_{1/2})_0]+A_3\Phi[(2p_{3/2}2p_{3/2})_0]+\ldots\,$,
         $\Psi[2s2p\,^3P_1]=B_1\Phi[(2s2p_{1/2})_1]+B_2\Phi[(2s2p_{3/2})_1]+\ldots\,$, and
         $\Psi[2s2p\,^1P_1]=C_1\Phi[(2s2p_{1/2})_1]+C_2\Phi[(2s2p_{3/2})_1]+\ldots\,$.
         The coefficients are obtained by means of the configuration-interaction method for a given configuration space. 
         The decomposition into the positive- and negative-energy spectra is determined by the Dirac Hamiltonian with the local Dirac-Fock potential included.
         }
         
\begin{tabular}{
                l@{\qquad}
                S[table-format=2.4]                
                S[table-format=2.4] 
                S[table-format=2.4]@{\qquad} 
                S[table-format=2.4] 
                S[table-format=-2.4]@{\qquad} 
                S[table-format=2.4] 
                S[table-format=2.4]
               }
               
\hline
\hline

   \multirow{2}{*}{$Z$}    &
   \multicolumn{3}{c}{\rule{0pt}{1.2em} $2s2s\,^1S_0$~~~~~~}                             &
   \multicolumn{2}{c}{\rule{0pt}{1.2em} $2s2p\,^3P_1$~~~~~~}                              &
   \multicolumn{2}{c}{\rule{0pt}{1.2em} $2s2p\,^1P_1$}                              \\
                                                         &
   \multicolumn{1}{c}{\rule[-0.4em]{0pt}{0.4em} $A_1$}                                                         &
   \multicolumn{1}{c}{\rule[-0.4em]{0pt}{0.4em} $A_2$}                                                          &
   \multicolumn{1}{c}{\rule[-0.4em]{0pt}{0.4em} $A_3$~~~~~~}                                                          &
   \multicolumn{1}{c}{\rule[-0.4em]{0pt}{0.4em} $B_1$}                                                          &
   \multicolumn{1}{c}{\rule[-0.4em]{0pt}{0.4em} $B_2$~~~~~~}                                                          &
   \multicolumn{1}{c}{\rule[-0.4em]{0pt}{0.4em} $C_1$}                                                          &
   \multicolumn{1}{c}{\rule[-0.4em]{0pt}{0.4em} $C_2$}                                                           \\ 
        
\hline   
                       
  18 \rule{0pt}{3.2ex}  &     0.9739 &     0.1371 &     0.1800 &     0.8461 &    -0.5326 &     0.5321 &     0.8454    \\ 
  36  &     0.9840 &     0.1378 &     0.1126 &     0.9602 &    -0.2790 &     0.2789 &     0.9601    \\ 
  42  &     0.9864 &     0.1375 &     0.0897 &     0.9799 &    -0.1992 &     0.1991 &     0.9798    \\ 
  54  &     0.9895 &     0.1337 &     0.0550 &     0.9948 &    -0.1016 &     0.1015 &     0.9947    \\ 
  82  &     0.9933 &     0.1137 &     0.0188 &     0.9996 &    -0.0275 &     0.0275 &     0.9996    \\ 
  83  &     0.9934 &     0.1129 &     0.0182 &     0.9996 &    -0.0264 &     0.0264 &     0.9996    \\ 
  90  &     0.9940 &     0.1081 &     0.0143 &     0.9998 &    -0.0200 &     0.0200 &     0.9998    \\ 
  92  &     0.9942 &     0.1070 &     0.0133 &     0.9998 &    -0.0186 &     0.0185 &     0.9998    \\ 

\hline
\hline

\end{tabular}%

\end{table*}

In this section, we present the results of our QED calculations for Be-like HCIs. First, we demonstrate the level mixing using the results of the CI method. Second, we provide an error budget and estimate the uncertainties associated with the uncalculated higher-order contributions. Finally, we present our theoretical predictions for the $2s2s \, ^1S_0 \rightarrow 2s2p \, ^3P_{0,1,2}$ and $2s2s \, ^1S_0 \rightarrow 2s2p \, ^1P_1$ excitation energies and for the $2s2p \, ^3P_1 \rightarrow 2s2p \, ^3P_0$ and $2s2p \, ^3P_2 \rightarrow 2s2p \, ^3P_1$ transition energies. These predictions are compared with available experimental data and results of previous relativistic calculations. 

Since the CI method is a non-perturbative approach with respect to the interelectronic interaction, the level mixing does not pose a problem in its framework. In our implementation of the CI method, the DCB operator acts in a space of Slater determinants constructed from the positive-energy eigenfunctions of the Dirac Hamiltonian. The latter includes the same local binding potential as that which determines the QED PT. In principle, other one-electron operators, e.g., the nonlocal Dirac-Fock one, can be used to determine a decomposition into the positive- and negative-energy spectra. There is an ambiguity in this matter within the Breit approximation~\cite{Sapirstein:1999:259}. We use the variant specified above, since it follows naturally from QED~\cite{Shabaev:1993:4703, Shabaev:2024:94:inbook} and thus provides the most accurate merging with the QED calculations. Nevertheless, the ambiguity of the Breit-approximation results associated with the choice of the local binding potential still remains, and it can be resolved only within the rigorous QED approach, see the related discussions in Refs.~\cite{Kozhedub:2019:062506, Malyshev:2023:042806}.

In Table~\ref{tab:amplitude}, an example of our CI calculations of Be-like HCIs is presented. Within the CI method, the many-electron wave functions are sought in the form of an expansion in configuration-state functions (CSFs) with the given values of the total angular momentum~$J$ and its projection~$M_J$. Table~\ref{tab:amplitude} shows the coefficients for the CSFs assigned to the model subspaces described above. Namely, the coefficients $A_1$, $A_2$, and $A_3$ correspond to the $(2s2s)_0$, $(2p_{1/2}2p_{1/2})_0$, and $(2p_{3/2}2p_{3/2})_0$ levels contributing to the ground state. The coefficients $B_{1}$ and $B_{2}$ correspond to the $(2s2p_{1/2})_1$ and $(2s2p_{3/2})_1$ levels in the $2s2p \, ^3P_1$ state, while the coefficients $C_{1}$ and $C_{2}$ have the same meaning for the $2s2p \, ^1P_1$ state. The values of the coefficients presented in Table~\ref{tab:amplitude} are not to be regarded as the exact or final ones. They are obtained for a certain finite size of configuration space, prior to any extrapolation to the infinite-dimensional space, and serve only to illustrate the level mixing. Moreover, the coefficients in Table~\ref{tab:amplitude} correspond to the Dirac Hamiltonian with the LDF potential included. For other potentials, the values will also change slightly. Nevertheless, the presented coefficients are found in good agreement with the ones given in Ref.~\cite{Armstrong:1976:1114}. 

The weights of the CSFs are determined by the squares of the coefficients. We note that in Table~\ref{tab:amplitude} for each state the sum of the coefficient squares is slightly less than one. The difference determines the contribution of CSFs lying beyond the model subspaces. From Table~\ref{tab:amplitude} it is seen that for all the states the coefficients corresponding to the leading configurations, $A_1$, $B_1$, and $C_2$, tend to one, as the nuclear charge $Z$ increases. The quasidegeneracy is gradually lifted with the growth of $Z$. Nevertheless, even for $Z=92$ the admixture of the $(2p_{1/2}2p_{1/2})_0$ CSF to the ground state remain significant. It is noteworthy that for low-$Z$ Be-like HCIs the contribution of the $(2p_{3/2}2p_{3/2})_0$ level to the ground state exceeds that of the $(2p_{1/2}2p_{1/2})_0$ level. The latter indicates that it is indeed crucial to consider the three-dimensional model subspace. We also note that the coefficients of the $(2s2p_{1/2})_1$ and $(2s2p_{3/2})_1$ levels for the $2s2p \, ^3P_1$ and $2s2p \, ^1P_1$ states seem to tend to the $jj$-$LS$ recoupling matrix for these pairs of states, see, e.g., Refs.~\cite{Drake:1988:586, Artemyev:2005:062104} for the related application of this matrix to the QED calculations. This behavior is consistent with the well-known fact that the $LS$ coupling is more appropriate for describing light atoms than the $jj$ coupling. Finally, we stress that for the $2s2p \, ^3P_0$ and $2s2p \, ^3P_2$ states in Be-like argon the coefficients of the leading levels $(2s2p_{1/2})_0$ and $(2s2p_{3/2})_2$ are both equal to $0.9998$ and become even closer to one as $Z$ increases. This justifies the use of PT for a single level to describe these states.

\onecolumngrid
{
\renewcommand{\arraystretch}{1.25}

\begin{longtable}{
                  l
                  S[table-format=1.6, group-separator=, table-space-text-pre=$<\,$]                
                  S[table-format=1.6, group-separator=, table-space-text-pre=$<\,$] 
                  S[table-format=1.6, group-separator=, table-space-text-pre=$<\,$] 
                  S[table-format=1.6, group-separator=, table-space-text-pre=$<\,$] 
                  S[table-format=1.6, group-separator=, table-space-text-pre=$<\,$] 
                  S[table-format=1.6, group-separator=, table-space-text-pre=$<\,$] 
                 }

\caption{\label{tab:error} 
         Error budget for the excitation and transition energies (in eV).
         See the text for details.
         }\\
               
\toprule
\toprule

   \multirow{2}{*}{\parbox{2.8cm}{\centering Source}}    &
   \multicolumn{1}{c}{\parbox{1.8cm}{\centering \rule{0pt}{1.2em} $2s2p\,^3P_0$}}                             &
   \multicolumn{1}{c}{\parbox{1.8cm}{\centering \rule{0pt}{1.2em} $2s2p\,^3P_1$}}                              &
   \multicolumn{1}{c}{\parbox{1.8cm}{\centering \rule{0pt}{1.2em} $2s2p\,^3P_2$}}                              &
   \multicolumn{1}{c}{\parbox{1.8cm}{\centering \rule{0pt}{1.2em} $2s2p\,^1P_1$}}                              &
   \multicolumn{1}{c}{\parbox{1.8cm}{\centering \rule{0pt}{1.2em} $2s2p\,^3P_1$}}                              &
   \multicolumn{1}{c}{\parbox{1.8cm}{\centering \rule{0pt}{1.2em} $2s2p\,^3P_2$}}                               \\
                                                         &
   \multicolumn{1}{c}{\parbox{1.8cm}{\centering \rule[-0.4em]{0pt}{0.4em} $- 2s2s\,^1S_0$}}                                                         &
   \multicolumn{1}{c}{\parbox{1.8cm}{\centering \rule[-0.4em]{0pt}{0.4em} $- 2s2s\,^1S_0$}}                                                          &
   \multicolumn{1}{c}{\parbox{1.8cm}{\centering \rule[-0.4em]{0pt}{0.4em} $- 2s2s\,^1S_0$}}                                                          &
   \multicolumn{1}{c}{\parbox{1.8cm}{\centering \rule[-0.4em]{0pt}{0.4em} $- 2s2s\,^1S_0$}}                                                          &
   \multicolumn{1}{c}{\parbox{1.8cm}{\centering \rule[-0.4em]{0pt}{0.4em} $- 2s2p\,^3P_0$}}                                                          &
   \multicolumn{1}{c}{\parbox{1.8cm}{\centering \rule[-0.4em]{0pt}{0.4em} $- 2s2p\,^3P_1$}}                                                           \\ 
        
\midrule

\endfirsthead

\caption[]{Error budget \textit{(continued)}.}\\

\toprule
\toprule

   \multirow{2}{*}{\parbox{2.8cm}{\centering Source}}    &
   \multicolumn{1}{c}{\parbox{1.8cm}{\centering \rule{0pt}{1.2em} $2s2p\,^3P_0$}}                             &
   \multicolumn{1}{c}{\parbox{1.8cm}{\centering \rule{0pt}{1.2em} $2s2p\,^3P_1$}}                              &
   \multicolumn{1}{c}{\parbox{1.8cm}{\centering \rule{0pt}{1.2em} $2s2p\,^3P_2$}}                              &
   \multicolumn{1}{c}{\parbox{1.8cm}{\centering \rule{0pt}{1.2em} $2s2p\,^1P_1$}}                              &
   \multicolumn{1}{c}{\parbox{1.8cm}{\centering \rule{0pt}{1.2em} $2s2p\,^3P_1$}}                              &
   \multicolumn{1}{c}{\parbox{1.8cm}{\centering \rule{0pt}{1.2em} $2s2p\,^3P_2$}}                               \\
                                                         &
   \multicolumn{1}{c}{\parbox{1.8cm}{\centering \rule[-0.4em]{0pt}{0.4em} $- 2s2s\,^1S_0$}}                   &
   \multicolumn{1}{c}{\parbox{1.8cm}{\centering \rule[-0.4em]{0pt}{0.4em} $- 2s2s\,^1S_0$}}                    &
   \multicolumn{1}{c}{\parbox{1.8cm}{\centering \rule[-0.4em]{0pt}{0.4em} $- 2s2s\,^1S_0$}}                    &
   \multicolumn{1}{c}{\parbox{1.8cm}{\centering \rule[-0.4em]{0pt}{0.4em} $- 2s2s\,^1S_0$}}                    &
   \multicolumn{1}{c}{\parbox{1.8cm}{\centering \rule[-0.4em]{0pt}{0.4em} $- 2s2p\,^3P_0$}}                    &
   \multicolumn{1}{c}{\parbox{1.8cm}{\centering \rule[-0.4em]{0pt}{0.4em} $- 2s2p\,^3P_1$}}                     \\ 
       
\midrule  

\endhead
\endfoot
\bottomrule
\bottomrule
\endlastfoot
       
  \multicolumn{7}{c}{\rule{0pt}{3.1ex} $Z=18$}   \\ 
 1el FNS                 &    0.000001  &    0.000001  &    0.000001  &    0.000001  &  {$<\,$}0.000001  &  {$<\,$}0.000001    \\ 
 1el nucl. pol.          &    0.000001  &    0.000001  &    0.000001  &    0.000001  &  {$<\,$}0.000001  &  {$<\,$}0.000001    \\ 
 1el 2loop               &    0.000005  &    0.000005  &    0.000005  &    0.000005  &  {$<\,$}0.000001  &    0.000003    \\ 
 H.o. QED, $\geqslant$3ph &     0.00024  &     0.00024  &     0.00024  &     0.00024  &     0.00024  &     0.00024    \\ 
 H.o. QED, 2loop         &    0.000088  &    0.000088  &    0.000088  &    0.000088  &    0.000088  &    0.000088    \\ 
 H.o. QED, scatter       &     0.00027  &     0.00046  &     0.00024  &     0.00058  &     0.00019  &     0.00025    \\ 
 Other                   &     0.00019  &     0.00027  &     0.00019  &     0.00029  &     0.00023  &     0.00024    \\ 
 Total \rule[-0.6em]{0pt}{0.6em}    &     0.00041  &     0.00059  &     0.00039  &     0.00070  &     0.00039  &     0.00043    \\ 
  \multicolumn{7}{c}{\rule{0pt}{3.1ex} $Z=36$}   \\ 
 1el FNS                 &     0.00005  &     0.00005  &     0.00005  &     0.00005  &  {$<\,$}0.00001  &  {$<\,$}0.00001    \\ 
 1el nucl. pol.          &     0.00001  &     0.00001  &     0.00001  &     0.00001  &  {$<\,$}0.00001  &  {$<\,$}0.00001    \\ 
 1el 2loop               &     0.00025  &     0.00025  &     0.00026  &     0.00026  &  {$<\,$}0.00001  &     0.00011    \\ 
 H.o. QED, $\geqslant$3ph &      0.0014  &      0.0014  &      0.0014  &      0.0014  &      0.0014  &      0.0014    \\ 
 H.o. QED, 2loop         &      0.0011  &      0.0011  &      0.0011  &      0.0011  &      0.0011  &      0.0011    \\ 
 H.o. QED, scatter       &     0.00094  &     0.00086  &     0.00028  &     0.00094  &     0.00008  &     0.00058    \\ 
 Other                   &     0.00030  &     0.00033  &     0.00030  &     0.00041  &     0.00030  &     0.00030    \\ 
 Total \rule[-0.6em]{0pt}{0.6em}    &      0.0021  &      0.0020  &      0.0019  &      0.0021  &      0.0018  &      0.0019    \\ 
  \multicolumn{7}{c}{\rule{0pt}{3.1ex} $Z=42$}   \\ 
 1el FNS                 &     0.00011  &     0.00011  &     0.00011  &     0.00011  &  {$<\,$}0.00001  &  {$<\,$}0.00001    \\ 
 1el nucl. pol.          &     0.00010  &     0.00010  &     0.00010  &     0.00010  &  {$<\,$}0.00001  &  {$<\,$}0.00001    \\ 
 1el 2loop               &     0.00061  &     0.00061  &     0.00063  &     0.00063  &  {$<\,$}0.00001  &     0.00028    \\ 
 H.o. QED, $\geqslant$3ph &      0.0022  &      0.0022  &      0.0022  &      0.0022  &      0.0022  &      0.0022    \\ 
 H.o. QED, 2loop         &      0.0018  &      0.0018  &      0.0018  &      0.0018  &      0.0018  &      0.0018    \\ 
 H.o. QED, scatter       &      0.0012  &      0.0012  &     0.00007  &     0.00069  &     0.00007  &      0.0013    \\ 
 Other                   &     0.00043  &     0.00046  &     0.00043  &     0.00050  &     0.00042  &     0.00042    \\ 
 Total \rule[-0.6em]{0pt}{0.6em}    &      0.0032  &      0.0032  &      0.0029  &      0.0030  &      0.0029  &      0.0031    \\ 
  \multicolumn{7}{c}{\rule{0pt}{3.1ex} $Z=54$}   \\ 
 1el FNS                 &     0.00097  &     0.00097  &     0.00097  &     0.00097  &  {$<\,$}0.00001  &     0.00003    \\ 
 1el nucl. pol.          &     0.00047  &     0.00047  &     0.00047  &     0.00047  &  {$<\,$}0.00001  &     0.00001    \\ 
 1el 2loop               &      0.0027  &      0.0027  &      0.0028  &      0.0028  &  {$<\,$}0.00001  &      0.0012    \\ 
 H.o. QED, $\geqslant$3ph &      0.0044  &      0.0044  &      0.0044  &      0.0044  &      0.0044  &      0.0044    \\ 
 H.o. QED, 2loop         &      0.0043  &      0.0043  &      0.0043  &      0.0043  &      0.0043  &      0.0043    \\ 
 H.o. QED, scatter       &      0.0026  &      0.0024  &      0.0012  &     0.00030  &     0.00030  &      0.0037    \\ 
 Other                   &     0.00082  &     0.00083  &     0.00080  &     0.00086  &     0.00080  &     0.00078    \\ 
 Total \rule[-0.6em]{0pt}{0.6em}    &      0.0074  &      0.0073  &      0.0070  &      0.0069  &      0.0063  &      0.0073    \\ 
  \multicolumn{7}{c}{\rule{0pt}{3.1ex} $Z=82$}   \\ 
 1el FNS                 &       0.022  &       0.022  &       0.022  &       0.022  &  {$<\,$}0.0001  &      0.0017    \\ 
 1el nucl. pol.          &      0.0025  &      0.0025  &      0.0025  &      0.0025  &  {$<\,$}0.0001  &      0.0002    \\ 
 1el 2loop               &       0.038  &       0.038  &       0.037  &       0.037  &  {$<\,$}0.0001  &      0.0090    \\ 
 H.o. QED, $\geqslant$3ph &       0.017  &       0.017  &       0.017  &       0.017  &       0.017  &       0.017    \\ 
 H.o. QED, 2loop         &       0.019  &       0.019  &       0.019  &       0.019  &       0.019  &       0.019    \\ 
 H.o. QED, scatter       &       0.015  &       0.013  &      0.0084  &      0.0056  &      0.0025  &       0.021    \\ 
 Other                   &      0.0036  &      0.0036  &      0.0035  &      0.0036  &      0.0031  &      0.0032    \\ 
 Total \rule[-0.6em]{0pt}{0.6em}    &       0.053  &       0.052  &       0.051  &       0.051  &       0.026  &       0.035    \\ 
  \multicolumn{7}{c}{\rule{0pt}{3.1ex} $Z=83$}   \\ 
 1el FNS                 &       0.026  &       0.026  &       0.026  &       0.026  &  {$<\,$}0.0001  &      0.0021    \\ 
 1el nucl. pol.          &       0.013  &       0.013  &       0.013  &       0.013  &  {$<\,$}0.0001  &      0.0012    \\ 
 1el 2loop               &       0.041  &       0.041  &       0.040  &       0.040  &  {$<\,$}0.0001  &      0.0074    \\ 
 H.o. QED, $\geqslant$3ph &       0.018  &       0.018  &       0.018  &       0.018  &       0.018  &       0.018    \\ 
 H.o. QED, 2loop         &       0.020  &       0.020  &       0.020  &       0.020  &       0.020  &       0.020    \\ 
 H.o. QED, scatter       &       0.016  &       0.013  &      0.0087  &      0.0059  &      0.0027  &       0.022    \\ 
 Other                   &      0.0042  &      0.0042  &      0.0042  &      0.0042  &      0.0037  &      0.0038    \\ 
 Total \rule[-0.6em]{0pt}{0.6em}    &       0.059  &       0.058  &       0.057  &       0.057  &       0.027  &       0.036    \\ 
  \multicolumn{7}{c}{\rule{0pt}{3.1ex} $Z=90$}   \\ 
 1el FNS                 &        0.11  &        0.11  &        0.11  &        0.11  &  {$<\,$}0.0001  &       0.011    \\ 
 1el nucl. pol.          &       0.012  &       0.012  &       0.012  &       0.012  &  {$<\,$}0.0001  &      0.0012    \\ 
 1el 2loop               &       0.074  &       0.074  &       0.073  &       0.073  &  {$<\,$}0.0001  &       0.015    \\ 
 H.o. QED, $\geqslant$3ph &       0.025  &       0.025  &       0.025  &       0.025  &       0.025  &       0.025    \\ 
 H.o. QED, 2loop         &       0.027  &       0.027  &       0.027  &       0.027  &       0.027  &       0.027    \\ 
 H.o. QED, scatter       &       0.024  &       0.020  &       0.011  &       0.008  &      0.0040  &       0.031    \\ 
 Other                   &       0.010  &       0.010  &       0.009  &       0.009  &      0.0055  &      0.0062    \\ 
 Total \rule[-0.6em]{0pt}{0.6em}    &        0.14  &        0.14  &        0.13  &        0.13  &       0.037  &       0.052    \\ 
  \multicolumn{7}{c}{\rule{0pt}{3.1ex} $Z=92$}   \\ 
 1el FNS                 &       0.034  &       0.034  &       0.034  &       0.034  &  {$<\,$}0.0001  &      0.0040    \\ 
 1el nucl. pol.          &       0.020  &       0.020  &       0.020  &       0.020  &  {$<\,$}0.0001  &      0.0022    \\ 
 1el 2loop               &       0.088  &       0.088  &       0.086  &       0.086  &  {$<\,$}0.0001  &       0.018    \\ 
 H.o. QED, $\geqslant$3ph &       0.027  &       0.027  &       0.027  &       0.027  &       0.027  &       0.027    \\ 
 H.o. QED, 2loop         &       0.030  &       0.030  &       0.030  &       0.030  &       0.030  &       0.030    \\ 
 H.o. QED, scatter       &       0.026  &       0.022  &       0.012  &       0.008  &      0.0043  &       0.034    \\ 
 Other                   &       0.009  &       0.009  &       0.009  &       0.009  &      0.0068  &      0.0071    \\ 
 Total \rule[-0.6em]{0pt}{0.6em}    &        0.11  &        0.11  &        0.10  &        0.10  &       0.041  &       0.056    \\ 

\end{longtable}

}
\twocolumngrid

An essential part of the present study consists in a thorough analysis of the uncertainties that have to be ascribed to the theoretical predictions obtained. The error budget of our calculations is summarized in Table~\ref{tab:error}. The undertaken analysis of the uncertainties follows, with minor modifications, the scheme described in detail in Refs.~\cite{Malyshev:2021:183001, Malyshev:2023:042806}. For the sake of self-sufficiency, let us briefly discuss this issue here.

In Table~\ref{tab:error}, the first line ``1el FNS'' presents the uncertainties caused by the finite-nuclear-size (FNS) effect on the one-electron Dirac energies. For uranium, these uncertainties are obtained within the evaluation of the nuclear-deformation correction~\cite{Kozhedub:2008:032501}. For other ions, they are estimated by quadratically summing the variation due to changing the root-mean-square radius within its error bars and the variation due to replacing the Fermi model of the nuclear-charge distribution by a homogeneously-charged-sphere one. The second line ``1el nucl. pol.'' shows the uncertainties associated with the nuclear-polarization effect. They are estimated according to a prescription formulated in Ref.~\cite{Yerokhin:2015:033103} for H-like ions. The third line ``1el 2loop'' gives the uncertainties of the one-electron two-loop contributions. They are also taken from Ref.~\cite{Yerokhin:2015:033103}. Because of the level mixing, these one-electron uncertainties do not vanish even for the $2s2p \, ^3P_1 \rightarrow 2s2p \, ^3P_0$ transition energy. The three subsequent lines in Table~\ref{tab:error} provide the results of our various estimations of the uncalculated higher-order QED effects. First, in the line ``H.o. QED, $\geqslant$3ph'', we estimate the QED correction to the interelectronic-interaction contribution related to the exchange by three or more photons. For all the excitation and transition energies, we assign the same value of this uncertainty according to the procedure from Ref.~\cite{Kozhedub:2019:062506}. Namely, the third- and higher-order Breit-approximation value obtained within the CI method is multiplied by the doubled ratio of the corresponding second-order QED correction and its Breit-approximation counterpart. The ground-state binding energy evaluated for the LDF potential is used for this purpose. The calculations show that for other states the procedure gives values of the same order of magnitude. Second, in the line ``H.o. QED, 2loop'', we estimate the screening of the one-electron two-loop contribution by multiplying the corresponding term for the $1s$ state by a conservative factor of $2/Z$. Third, in the line ``H.o. QED, scatter'', we estimate the convergence of the perturbation series by analyzing the scatter of the results obtained for different binding potentials~$V$. We take the absolute value of the difference between the total results for the LDF and CH potentials and add to it the absolute value of the correction obtained for the LDF potential using the model-QED-operator approach. The line labeled ``Other'' summarizes all other sources of the uncertainties: (i) numerical errors of the QED and CI calculations, which are estimated by analyzing the convergence of the results with respect to the partial-wave expansions and the sizes of the finite-basis sets used; (ii) uncertainties due to the remaining part of the FNS effect, including that for the nuclear recoil contribution~\cite{Shabaev:1998:4235, Aleksandrov:2015:144004, Anisimova:2022:062823, Pachucki:2023:053002}; (iii) an estimate of the QED part of the nuclear recoil effect beyond the independent-electron approximation; and (iv) a similar estimate for the nuclear-polarization effect. The latter two estimates are obtained by multiplying the corresponding one-electron contributions for the $1s$ state by a factor of $2/Z$. The consideration of this additional uncertainty for the nuclear-polarization effect compared to the one given in the row~``1el nucl. pol.'' is obviously meaningless for the $2s2s \, ^1S_0 \rightarrow 2s2p \, ^3P_{0,1,2}$ and $2s2s \, ^1S_0 \rightarrow 2s2p \, ^1P_1$ excitation energies. However, this has sense for the $2s2p \, ^3P_1 \rightarrow 2s2p \, ^3P_0$ and $2s2p \, ^3P_2 \rightarrow 2s2p \, ^3P_1$ transition energies. The total uncertainties are evaluated by quadratically summing all the errors listed above. The convergence of the perturbation series and the overall performance of the numerical methods are kept under control by carrying out the calculations for the Coulomb, CH, and LDF potentials. For the same purposes, the calculations are performed in both Feynman and Coulomb gauge for the photons mediating the interelectronic interaction, see the related discussion, e.g., in Ref.~\cite{Malyshev:2023:042806}. An example of how perturbation series starting from different initial approximations converge can be found in Refs.~\cite{Malyshev:2017:022512, Malyshev:2023:042806}.

In order to better represent the competing sources of theoretical uncertainty, the individual contributions in Table~\ref{tab:error} are given with extra significant digits. To emphasize that some source provides a small but nonzero value, we indicate an inequality, such as ``$< 0.000001$''. From Table~\ref{tab:error}, it can be seen that for low-$Z$ Be-like HCIs the total uncertainties arise mainly from the uncalculated higher-order QED effects. For the $2s2p \, ^3P_1 \rightarrow 2s2p \, ^3P_0$ and $2s2p \, ^3P_2 \rightarrow 2s2p \, ^3P_1$ transition energies, this holds true also for high-$Z$ ions. However, for the $2s2s \, ^1S_0 \rightarrow 2s2p \, ^3P_{0,1,2}$ and $2s2s \, ^1S_0 \rightarrow 2s2p \, ^1P_1$ excitation energies in heavy Be-like HCIs, the situation currently differs. In this case, the total theoretical uncertainties are determined by the calculations of the one-electron two-loop contribution. The uncertainties due to the FNS and nuclear-polarization effects also come into play.

\onecolumngrid
{
\renewcommand{\arraystretch}{1.25}

\begin{longtable}{
                  S[table-format=3.5(2),group-separator=]                
                  S[table-format=3.5(2),group-separator=] 
                  S[table-format=3.5(2),group-separator=] 
                  S[table-format=3.5(2),group-separator=] 
                  l@{\quad}
                  l@{\quad}
                  l
                 }

\caption{\label{tab:excitation} 
         The excitation energies of the $2s2p\,^3P_{0,1,2}$ and $2s2p\,^1P_1$ states from the $2s2s\,^1S_0$ ground state in Be-like ions (in eV).
         The theoretical (Th.) results are compared with the experimental (Expt.) values.
         }\\
               
\toprule
\toprule

   \multicolumn{1}{c}{\parbox{1.8cm}{\centering \rule{0pt}{1.2em} $2s2p\,^3P_0$}~~~~~~}                             &
   \multicolumn{1}{c}{\parbox{1.8cm}{\centering \rule{0pt}{1.2em} $2s2p\,^3P_1$}~~~~~~}                              &
   \multicolumn{1}{c}{\parbox{1.8cm}{\centering \rule{0pt}{1.2em} $2s2p\,^3P_2$}~~~~~~}                              &
   \multicolumn{1}{c}{\parbox{1.8cm}{\centering \rule{0pt}{1.2em} $2s2p\,^1P_1$}~~~~~~}                              &
   \multirow{2}{*}{Th./Expt.}                                                                                    &
   \multirow{2}{*}{Year}                                                                                         &
   \multirow{2}{*}{Reference}                                                                                    \\                                                                                

   \multicolumn{1}{c}{\parbox{1.8cm}{\centering \rule[-0.4em]{0pt}{0.4em} $- 2s2s\,^1S_0$}~~~~~~}                   &
   \multicolumn{1}{c}{\parbox{1.8cm}{\centering \rule[-0.4em]{0pt}{0.4em} $- 2s2s\,^1S_0$}~~~~~~}                    &
   \multicolumn{1}{c}{\parbox{1.8cm}{\centering \rule[-0.4em]{0pt}{0.4em} $- 2s2s\,^1S_0$}~~~~~~}                    &
   \multicolumn{1}{c}{\parbox{1.8cm}{\centering \rule[-0.4em]{0pt}{0.4em} $- 2s2s\,^1S_0$}~~~~~~}                    &
                                                                                                                 &
                                                                                                                 &
                                                                                                                 \\
        
\midrule

\endfirsthead

\caption[]{Excitation energies \textit{(continued)}.}\\

\toprule
\toprule

   \multicolumn{1}{c}{\parbox{1.8cm}{\centering \rule{0pt}{1.2em} $2s2p\,^3P_0$}~~~~~~}                             &
   \multicolumn{1}{c}{\parbox{1.8cm}{\centering \rule{0pt}{1.2em} $2s2p\,^3P_1$}~~~~~~}                              &
   \multicolumn{1}{c}{\parbox{1.8cm}{\centering \rule{0pt}{1.2em} $2s2p\,^3P_2$}~~~~~~}                              &
   \multicolumn{1}{c}{\parbox{1.8cm}{\centering \rule{0pt}{1.2em} $2s2p\,^1P_1$}~~~~~~}                              &
   \multirow{2}{*}{Th./Expt.}                                                                                    &
   \multirow{2}{*}{Year}                                                                                         &
   \multirow{2}{*}{Reference}                                                                                    \\                                                                                

   \multicolumn{1}{c}{\parbox{1.8cm}{\centering \rule[-0.4em]{0pt}{0.4em} $- 2s2s\,^1S_0$}~~~~~~}                   &
   \multicolumn{1}{c}{\parbox{1.8cm}{\centering \rule[-0.4em]{0pt}{0.4em} $- 2s2s\,^1S_0$}~~~~~~}                    &
   \multicolumn{1}{c}{\parbox{1.8cm}{\centering \rule[-0.4em]{0pt}{0.4em} $- 2s2s\,^1S_0$}~~~~~~}                    &
   \multicolumn{1}{c}{\parbox{1.8cm}{\centering \rule[-0.4em]{0pt}{0.4em} $- 2s2s\,^1S_0$}~~~~~~}                    &
                                                                                                                 &
                                                                                                                 &
                                                                                                                 \\
       
\midrule  

\endhead
\endfoot
\bottomrule
\bottomrule
\endlastfoot
       
  \multicolumn{7}{c}{\rule{0pt}{3.1ex} $Z=18$}   \\ 
        28.35403(41)  &        29.24427(59)  &        31.32954(39)  &        56.06799(70)  &   Th.     &  2024  &  This work    \\ 

    28.354(95) &     29.244(95)  &    31.331(95)   &    56.069(95)   &   Th.     &  2019  &  Kaygorodov \textit{et al.} \cite{Kaygorodov:2019:032505}      \\
    
    28.3489    &     29.2451     &    31.3201      &                 &   Th.     &  2017  &  Li \textit{et al.} \cite{Li:2017:720}                         \\
    
    28.360(12) &     29.248(16)  &    31.334(12)   &    56.085(29)   &   Th.     &  2015  &  Yerokhin \textit{et al.}  \cite{Yerokhin:2015:054003}         \\
    
    
    28.3673    &     29.2585     &    31.3451      &    56.0268      &   Th.     &  2005  &  Gu \cite{Gu:2005:267}                                         \\   

               &     29.243      &    31.330       &    55.990       &   Th.     &  2000  &  Safronova \cite{Safronova:2000:1213}                          \\   
               
               &                 &    31.3293      &                 &   Th.     &  2000  &  Majumder and Das \cite{Majumder:2000:042508}                  \\ 
    
    28.3489    &     29.2397     &    31.3268      &    55.9872      &   Th.     &  1996  &  Safronova \textit{et al.} \cite{Safronova:1996:4036}          \\
    
    28.3520    &     29.2433     &    31.3287      &    56.0671      & Th.$^\dagger$ & 1983 & Edl\'en \cite{Edlen:1983:51}                                 \\ 
    
  28.3532(48)  &    29.2429(20)  &  31.3283(20)    &  56.0634(81)    & Expt.$^\ddagger$ &  2010  &  Saloman \cite{Saloman:2010:033101}                     \\
  
               &    29.243(2)    &                 &  56.071(8)      &   Expt.   &  1978  &  Dere \cite{Dere:1978:1062}                                    \\
               
               &    29.241(2)    &                 &  56.063(8)      &   Expt.   &  1975  &  Widing \cite{Widing:1975:L33} \rule[-0.6em]{0pt}{0.6em}       \\

  \multicolumn{7}{c}{\rule{0pt}{3.1ex} $Z=36$}   \\ 
         62.6306(21)  &         72.9862(20)  &        125.6572(19)  &        170.4200(21)  &   Th.     &  2024  &  This work    \\ 

    62.64(73)  &     72.98(73)   &    125.69(73)   &    170.45(73)   &   Th.     &  2019  &  Kaygorodov \textit{et al.} \cite{Kaygorodov:2019:032505}      \\
    
    62.6228    &     72.9867     &    125.656      &    170.454      &   Th.     &  2008  &  Cheng \textit{et al.} \cite{Cheng:2008:052504}                \\
    
    62.6970    &     73.0647     &    125.7730     &    170.5499     &   Th.     &  2005  &  Gu \cite{Gu:2005:267}                                         \\           

               &     72.975      &    125.639      &    170.397      &   Th.     &  2000  &  Safronova \cite{Safronova:2000:1213}                          \\     
    
    62.6198    &     72.9751     &    125.6409     &    170.3990     &   Th.     &  1996  &  Safronova \textit{et al.} \cite{Safronova:1996:4036}          \\   
    
    62.5216    &     72.8834     &    125.6282     &    170.6618     & Th.$^\dagger$ & 1983 & Edl\'en \cite{Edlen:1983:51}                                 \\
               
    62.67(12)  &                 &                 &                 &   Expt.   &  1991  &  Sugar and Musgrove \cite{Sugar:1991:859}                      \\
    
               &     72.998(11)  &    125.651(16)  &    170.411(47)  &   Expt.   &  1989  &  Denne \textit{et al.} \cite{Denne:1989:1488} \rule[-0.6em]{0pt}{0.6em}   \\

  \multicolumn{7}{c}{\rule{0pt}{3.1ex} $Z=42$}   \\ 
         75.2876(32)  &         90.0049(32)  &        197.9864(29)  &        248.4990(30)  &   Th.     &  2024  &  This work    \\ 

    75.3(12)   &     90.0(12)    &    198.0(12)    &    248.5(12)    &   Th.     &  2019  &  Kaygorodov \textit{et al.} \cite{Kaygorodov:2019:032505}      \\
    
    75.2724    &     90.0053     &    197.982      &    248.540      &   Th.     &  2008  &  Cheng \textit{et al.} \cite{Cheng:2008:052504}                \\
    
    75.3798    &     90.1169     &    198.1860     &    248.7330     &   Th.     &  2005  &  Gu \cite{Gu:2005:267}                                         \\
    
               &     89.981      &    197.94       &    248.47       &   Th.     &  2000  &  Safronova \cite{Safronova:2000:1213}                          \\ 
    
               &     90.0061     &                 &    248.5428     &   Th.     &  1997  &  Chen and Cheng \cite{Chen:1997:166}                           \\    
               
    75.2691    &     89.9895     &    197.9554     &    248.4831     &   Th.     &  1996  &  Safronova \textit{et al.} \cite{Safronova:1996:4036}          \\   
               
               &     89.982(37)  &                 &    248.523(33)  &   Th.     &  1992  &  Lindroth and Hvarfner \cite{Lindroth:1992:2771}               \\
               
               &     89.983(20)  &                 &    248.45(15)   &   Expt.   &  1989  &  Denne \textit{et al.} \cite{Denne:1989:3702} \rule[-0.6em]{0pt}{0.6em}   \\

  \multicolumn{7}{c}{\rule{0pt}{3.1ex} $Z=54$}   \\ 
        104.5305(74)  &        127.2997(73)  &        469.4832(70)  &        532.8009(69)  &   Th.     &  2024  &  This work    \\ 

    104.5(25)  &     127.3(25)   &    469.6(25)    &    532.9(25)    &   Th.     &  2019  &  Kaygorodov \textit{et al.} \cite{Kaygorodov:2019:032505}      \\
    
    104.475    &     127.282     &    469.449      &    532.877      &   Th.     &  2008  &  Cheng \textit{et al.} \cite{Cheng:2008:052504}                \\
    
    104.663    &     127.475     &    470.004      &    533.401      &   Th.     &  2005  &  Gu \cite{Gu:2005:267}                                         \\
    
               &     127.168     &    469.25       &    532.62       &   Th.     &  2000  &  Safronova \cite{Safronova:2000:1213}                          \\ 
               
               &     127.301     &                 &    532.854      &   Th.     &  1997  &  Chen and Cheng \cite{Chen:1997:166}                           \\
    
    104.482    &     127.267     &    469.386      &    532.759      &   Th.     &  1996  &  Safronova \textit{et al.} \cite{Safronova:1996:4036}          \\ 
    
               &     127.269(46) &    469.474(81)  &    532.801(16)  &   Expt.   &  2015  &  Bernhardt \textit{et al.} \cite{Bernhardt:2015:144008}        \\
               
               &     127.260(26) &                 &                 &   Expt.   &  2005  &  Feili \textit{et al.} \cite{Feili:2005:48}                    \\
               
               &     127.255(12) &                 &                 &   Expt.   &  2003  &  Tr\"abert \textit{et al.} \cite{Trabert:2003:042501} \rule[-0.6em]{0pt}{0.6em}   \\

  \multicolumn{7}{c}{\rule{0pt}{3.1ex} $Z=82$}   \\ 
         208.073(53)  &         244.942(52)  &        2584.793(51)  &        2687.576(51)  &   Th.     &  2024  &  This work    \\ 

    207.756    &     244.734     &    2584.54      &    2687.59      &   Th.     &  2008  &  Cheng \textit{et al.} \cite{Cheng:2008:052504}                \\
   
               &     244.845     &                 &    2687.43      &   Th.     &  1997  &  Chen and Cheng \cite{Chen:1997:166} \rule[-0.6em]{0pt}{0.6em}   \\
    
  \multicolumn{7}{c}{\rule{0pt}{3.1ex} $Z=83$}   \\ 
         212.956(59)  &         250.188(58)  &        2728.852(57)  &        2833.347(57)  &   Th.     &  2024  &  This work    \\ 

    213(11)    &     250(11)     &    2729(11)     &    2834(11)     &   Th.     &  2019  &  Kaygorodov \textit{et al.} \cite{Kaygorodov:2019:032505}      \\
   
    212.599    &     249.943     &    2728.56      &    2833.32      &   Th.     &  2008  &  Cheng \textit{et al.} \cite{Cheng:2008:052504}                \\

               &     248.03      &    2727         &    2832         &   Th.     &  2000  &  Safronova \cite{Safronova:2000:1213}                          \\
               
               &                 &                 &    2833.04      &   Th.     &  1998  &  Santos \textit{et al.} \cite{Santos:1998:149}                 \\               
               
    212.953    &     250.242     &    2728.459     &    2833.154     &   Th.     &  1996  &  Safronova \textit{et al.} \cite{Safronova:1996:4036} \rule[-0.6em]{0pt}{0.6em}   \\
    
  \multicolumn{7}{c}{\rule{0pt}{3.1ex} $Z=90$}   \\ 
          248.00(14)  &          287.38(14)  &         3951.40(13)  &         4068.64(13)  &   Th.     &  2024  &  This work    \\ 

    248(12)    &     288(12)     &    3952(12)     &    4069(12)     &   Th.     &  2019  &  Kaygorodov \textit{et al.} \cite{Kaygorodov:2019:032505}      \\
    
    247.461    &     286.968     &    3950.97      &    4068.47      &   Th.     &  2008  &  Cheng \textit{et al.} \cite{Cheng:2008:052504}                \\
    
               &                 &                 &    4068.54      &   Th.     &  2000  &  Cheng \textit{et al.} \cite{Cheng:2000:054501}                \\   
  
               &     284.30      &    3950         &    4068         &   Th.     &  2000  &  Safronova \cite{Safronova:2000:1213}                          \\   
               
               &                 &                 &    4068.21      &   Th.     &  1998  &  Santos \textit{et al.} \cite{Santos:1998:149}                 \\
               
               &     287.150     &                 &    4068.19      &   Th.     &  1997  &  Chen and Cheng \cite{Chen:1997:166}                           \\               

    248.072    &     287.511     &    3950.874     &    4068.360     &   Th.     &  1996  &  Safronova \textit{et al.} \cite{Safronova:1996:4036}          \\
    
               &                 &                 &    4068.47(16)  &   Expt.   &  1995  &  Beiersdorfer \textit{et al.} \cite{Beiersdorfer:1995:2693} \rule[-0.6em]{0pt}{0.6em}   \\
    
  \multicolumn{7}{c}{\rule{0pt}{3.1ex} $Z=92$}   \\ 
          258.07(11)  &          297.91(11)  &         4380.63(10)  &         4501.77(10)  &   Th.     &  2024  &  This work    \\ 

    259(13)    &     299(13)     &    4382(13)     &    4503(13)     &   Th.     &  2019  &  Kaygorodov \textit{et al.} \cite{Kaygorodov:2019:032505}      \\
                                                                                                                                                          
    257.564    &     297.537     &    4380.27      &    4501.66      &   Th.     &  2008  &  Cheng \textit{et al.} \cite{Cheng:2008:052504}                \\ 
                                                                                                                                                          
               &                 &                 &    4501.73      &   Th.     &  2000  &  Cheng \textit{et al.} \cite{Cheng:2000:054501}                \\ 
                                                                                                                                                          
               &     294.45      &    4380         &    4500         &   Th.     &  2000  &  Safronova \cite{Safronova:2000:1213}                          \\ 
                                                                                                                                                          
               &                 &                 &    4501.54      &   Th.     &  1998  &  Santos \textit{et al.} \cite{Santos:1998:149}                 \\
                                                                                                                                                          
               &     297.744     &                 &    4501.36      &   Th.     &  1997  &  Chen and Cheng \cite{Chen:1997:166}                           \\
                                                                                                                                                          
    258.276    &     298.177     &    4380.198     &    4501.602     &   Th.     &  1996  &  Safronova \textit{et al.} \cite{Safronova:1996:4036}          \\
                                                                                                                                                          
               &                 &                 &    4501.59(21)  &   Expt.   &  2024  &  Loetzsch \textit{et al.} \cite{Loetzsch:2024:673}             \\
                              
               &     297.799(12) &                 &                 &   Expt.   &  2005  &  Beiersdorfer \textit{et al.} \cite{Beiersdorfer:2005:233003}  \\
               
               &                 &                 &    4501.72(27)  &   Expt.   &  1993  &  Beiersdorfer \textit{et al.} \cite{Beiersdorfer:1993:3939} \rule[-0.6em]{0pt}{0.6em}   \\

\end{longtable}

$^\dagger$ Semiempirical prediction.

$^\ddagger$ Compilation of energy levels obtained by fitting to available lines.

\vspace*{2mm}

}
\twocolumngrid

Our theoretical predictions for the excitation energies of the $2s2p \, ^3P_{0,1,2}$ and $2s2p \, ^1P_1$ states from the $2s2s \, ^1S_0$ ground state and for the $2s2p \, ^3P_1 \rightarrow 2s2p \, ^3P_0$ and $2s2p \, ^3P_2 \rightarrow 2s2p \, ^3P_1$ transition energies are compiled in Tables~\ref{tab:excitation} and \ref{tab:3P2-3P1}, respectively. As the final results, we use the values obtained for the LDF potential and for the maximal sizes of the model subspaces indicated in Sec.~\ref{sec:1}. The uncertainties of the theoretical predictions are estimated according to the procedure summarized in Table~\ref{tab:error}. 

Tables~\ref{tab:excitation} and \ref{tab:3P2-3P1} give a detailed comparison of the excitation and transition energies with the previous relativistic calculations and available measurements. We note that for the $2s2p \, ^3P_2 \rightarrow 2s2p \, ^3P_1$ transition in Be-like argon the experimental wavelengths in air are usually reported in literature. The conversion between the air and vacuum wavelengths is accomplished using the three-term formula for the refractive index of air, which is given in Eq.~(3) in Ref.~\cite{Peck:1972:958}. The same approach was employed, e.g., in Ref.~\cite{Saloman:2010:033101}. As can be seen, the previous theoretical studies of Be-like HCIs indeed exhibit a considerable scatter of the results. In general, all these values are in agreement with our predictions. When the uncertainties are specified~\cite{Sampaio:2013:014015, Yerokhin:2015:054003, Kaygorodov:2019:032505}, they are, in principle, consistent. However, our results are much more precise. In Ref.~\cite{Kaygorodov:2019:032505}, we evaluated the energies of Be-like HCIs in the framework of the CI method, taking into account the nuclear-recoil and QED effect by means of the mass-shift~\cite{Shabaev:1985:394, Shabaev:1988:107, Palmer:1987:5987} and model-QED~\cite{Shabaev:2013:012513, Shabaev:2015:175:2018:69:join_pr} operators. The uncertainties in that study have been estimated in a rather conservative manner. A comparison with the present work shows that the accuracy of the approach used in Ref.~\cite{Kaygorodov:2019:032505} is at least one order of magnitude higher.

\onecolumngrid
{
\renewcommand{\arraystretch}{1.25}

\begin{longtable}{
                  S[table-format=4.5(2),group-separator=]
                  S[table-format=4.7(2),group-separator=]@{\quad}
                  l@{\quad}
                  l@{\quad}
                  l
                 }

\caption{\label{tab:3P2-3P1} 
         The $2s2p\,^3P_{1} \rightarrow 2s2p\,^3P_{0}$ and $2s2p\,^3P_{2} \rightarrow 2s2p\,^3P_{1}$ transition energies in Be-like ions (in eV).
         The theoretical (Th.) results are compared with the experimental (Expt.) values.
         }\\
               
\toprule
\toprule

   \multicolumn{1}{c}{\parbox{1.8cm}{\centering \rule{0pt}{1.2em} $2s2p\,^3P_1$}~~~~~~}                             &
   \multicolumn{1}{c}{\parbox{1.8cm}{\centering \rule{0pt}{1.2em} $2s2p\,^3P_2$}~~~~~~}                             &
   \multirow{2}{*}{Th./Expt.}                                                                                    &
   \multirow{2}{*}{Year}                                                                                         &
   \multirow{2}{*}{Reference}                                                                                    \\                                                                                

   \multicolumn{1}{c}{\parbox{1.8cm}{\centering \rule[-0.4em]{0pt}{0.4em} $- 2s2p\,^3P_0$}~~~~~~}                   &
   \multicolumn{1}{c}{\parbox{1.8cm}{\centering \rule[-0.4em]{0pt}{0.4em} $- 2s2p\,^3P_1$}~~~~~~}                   &
                                                                                                                 &
                                                                                                                 &
                                                                                                                 \\
        
\midrule

\endfirsthead

\caption[]{Transition energies \textit{(continued)}.}\\

\toprule
\toprule

   \multicolumn{1}{c}{\parbox{1.8cm}{\centering \rule{0pt}{1.2em} $2s2p\,^3P_1$}~~~~~~}                             &
   \multicolumn{1}{c}{\parbox{1.8cm}{\centering \rule{0pt}{1.2em} $2s2p\,^3P_2$}~~~~~~}                             &
   \multirow{2}{*}{Th./Expt.}                                                                                    &
   \multirow{2}{*}{Year}                                                                                         &
   \multirow{2}{*}{Reference}                                                                                    \\                                                                                

   \multicolumn{1}{c}{\parbox{1.8cm}{\centering \rule[-0.4em]{0pt}{0.4em} $- 2s2p\,^3P_0$}~~~~~~}                   &
   \multicolumn{1}{c}{\parbox{1.8cm}{\centering \rule[-0.4em]{0pt}{0.4em} $- 2s2p\,^3P_1$}~~~~~~}                   &
                                                                                                                 &
                                                                                                                 &
                                                                                                                 \\
       
\midrule  

\endhead
\endfoot
\bottomrule
\bottomrule
\endlastfoot
       
  \multicolumn{5}{c}{\rule{0pt}{3.1ex} $Z=18$}   \\ 
         0.89024(39)  &         2.08527(43)  &   Th.     &  2024  &  This work    \\ 

   0.890(95)        &   2.087(95)        &   Th.     &  2019  &  Kaygorodov \textit{et al.} \cite{Kaygorodov:2019:032505}      \\
   
   0.888(20)        &   2.086(20)        &   Th.     &  2015  &  Yerokhin \textit{et al.}  \cite{Yerokhin:2015:054003}         \\
   
   
                    &   2.0859(11)       &   Th.     &  2006  &  Soria Orts \textit{et al.} \cite{SoriaOrts:2006:103002}       \\
   
   0.8912           &   2.0866           &   Th.     &  2005  &  Gu \cite{Gu:2005:267}                                         \\
   
                    &   2.0839           &   Th.     &  2003  &  Dragani\'c \textit{et al.} \cite{Draganic:2003:183001}        \\
   
   0.8902           &   2.0807           &   Th.     &  2001  &  Dong \textit{et al.} \cite{Dong:2001:294}                     \\
   
   0.8908           &   2.0871           &   Th.     &  1996  &  Safronova \textit{et al.} \cite{Safronova:1996:4036}          \\
   
                    &   2.0854           & Th.$^\dagger$ & 1983 & Edl\'en \cite{Edlen:1983:51}                                 \\ 
   
                    &   2.085339(7)      &   Expt.   &  2007  &  Katai \textit{et al.} \cite{Katai:2007:120}                   \\
   
                    &   2.0853362(7)     &   Expt.   &  2006  &  Soria Orts \textit{et al.} \cite{SoriaOrts:2006:103002}       \\
   
                    &   2.0853358(18)    &   Expt.   &  2003  &  Dragani\'c \textit{et al.} \cite{Draganic:2003:183001}        \\
   
                    &   2.085388(14)     &   Expt.   &  1997  &  Bieber \textit{et al.} \cite{Bieber:1997:64} \rule[-0.6em]{0pt}{0.6em}       \\

  \multicolumn{5}{c}{\rule{0pt}{3.1ex} $Z=36$}   \\ 
         10.3556(18)  &         52.6710(19)  &   Th.     &  2024  &  This work    \\ 

   10.34(73)        &   52.71(73)        &   Th.     &  2019  &  Kaygorodov \textit{et al.} \cite{Kaygorodov:2019:032505}      \\
   
   10.392(31)       &                    &   Th.     &  2013  &  Sampaio \textit{et al.} \cite{Sampaio:2013:014015}            \\
   
   10.4341          &                    &   Th.     &  2011  &  Winters \textit{et al.} \cite{Winters:2011:014013}            \\
   
   10.3639          &   52.669           &   Th.     &  2008  &  Cheng \textit{et al.} \cite{Cheng:2008:052504}                \\    
   
   10.3677          &   52.7083          &   Th.     &  2005  &  Gu \cite{Gu:2005:267}                                         \\
   
   10.3553          &   52.6658          &   Th.     &  1996  &  Safronova \textit{et al.} \cite{Safronova:1996:4036}          \\
   
   10.535           &                    &   Th.     &  1993  &  Marques \textit{et al.} \cite{Marques:1993:929}               \\
   
                    &   52.7449          & Th.$^\dagger$ & 1983 & Edl\'en \cite{Edlen:1983:51}                                 \\
   
   10.33(26)        &   52.82(32)        &   Expt.   &  1990  &  Martin \textit{et al.} \cite{Martin:1990:6570}                \\
   
                    &   52.652(11)       &   Expt.   &  1989  &  Denne \textit{et al.} \cite{Denne:1989:1488} \rule[-0.6em]{0pt}{0.6em}       \\

  \multicolumn{5}{c}{\rule{0pt}{3.1ex} $Z=42$}   \\ 
         14.7174(29)  &        107.9815(31)  &   Th.     &  2024  &  This work    \\ 

   14.7(12)         &   108.0(12)        &   Th.     &  2019  &  Kaygorodov \textit{et al.} \cite{Kaygorodov:2019:032505}      \\
   
   14.7329          &   107.977          &   Th.     &  2008  &  Cheng \textit{et al.} \cite{Cheng:2008:052504}                \\    
   
   14.7371          &   108.0691         &   Th.     &  2005  &  Gu \cite{Gu:2005:267}                                         \\
   
   14.7204          &   107.9659         &   Th.     &  1996  &  Safronova \textit{et al.} \cite{Safronova:1996:4036} \rule[-0.6em]{0pt}{0.6em}       \\

  \multicolumn{5}{c}{\rule{0pt}{3.1ex} $Z=54$}   \\ 
         22.7693(63)  &        342.1835(73)  &   Th.     &  2024  &  This work    \\ 

   22.8(25)         &   342.3(25)        &   Th.     &  2019  &  Kaygorodov \textit{et al.} \cite{Kaygorodov:2019:032505}      \\
   
   22.807           &   342.167          &   Th.     &  2008  &  Cheng \textit{et al.} \cite{Cheng:2008:052504}                \\    
   
   22.812           &   342.529          &   Th.     &  2005  &  Gu \cite{Gu:2005:267}                                         \\
   
   22.785           &   342.119          &   Th.     &  1996  &  Safronova \textit{et al.} \cite{Safronova:1996:4036}          \\
   
   23.191           &                    &   Th.     &  1993  &  Marques \textit{et al.} \cite{Marques:1993:929} \rule[-0.6em]{0pt}{0.6em}       \\

  \multicolumn{5}{c}{\rule{0pt}{3.1ex} $Z=82$}   \\ 
          36.869(26)  &        2339.851(35)  &   Th.     &  2024  &  This work    \\ 
   
   36.978           &   2339.81          &   Th.     &  2008  &  Cheng \textit{et al.} \cite{Cheng:2008:052504}                \\    
   
   37.419           &                    &   Th.     &  1993  &  Marques \textit{et al.} \cite{Marques:1993:929} \rule[-0.6em]{0pt}{0.6em}       \\

  \multicolumn{5}{c}{\rule{0pt}{3.1ex} $Z=83$}   \\ 
          37.232(27)  &        2478.664(36)  &   Th.     &  2024  &  This work    \\ 

   37(11)           &   2479(11)         &   Th.     &  2019  &  Kaygorodov \textit{et al.} \cite{Kaygorodov:2019:032505}      \\
   
   37.344           &   2478.62          &   Th.     &  2008  &  Cheng \textit{et al.} \cite{Cheng:2008:052504}                \\    
   
   37.289           &   2478.217         &   Th.     &  1996  &  Safronova \textit{et al.} \cite{Safronova:1996:4036} \rule[-0.6em]{0pt}{0.6em}       \\

  \multicolumn{5}{c}{\rule{0pt}{3.1ex} $Z=90$}   \\ 
          39.379(37)  &        3664.020(52)  &   Th.     &  2024  &  This work    \\ 

   39(12)           &   3663(12)         &   Th.     &  2019  &  Kaygorodov \textit{et al.} \cite{Kaygorodov:2019:032505}      \\
   
   39.507           &   3664.00          &   Th.     &  2008  &  Cheng \textit{et al.} \cite{Cheng:2008:052504}                \\    
   
   39.439           &   3663.363         &   Th.     &  1996  &  Safronova \textit{et al.} \cite{Safronova:1996:4036} \rule[-0.6em]{0pt}{0.6em}       \\

  \multicolumn{5}{c}{\rule{0pt}{3.1ex} $Z=92$}   \\ 
          39.841(41)  &        4082.725(56)  &   Th.     &  2024  &  This work    \\ 

   40(13)           &   4083(13)         &   Th.     &  2019  &  Kaygorodov \textit{et al.} \cite{Kaygorodov:2019:032505}      \\
   
   39.973           &   4082.73          &   Th.     &  2008  &  Cheng \textit{et al.} \cite{Cheng:2008:052504}                \\    
   
   39.901           &   4082.021         &   Th.     &  1996  &  Safronova \textit{et al.} \cite{Safronova:1996:4036}          \\
                    
                    &   4081.72(27)      &   Expt.   &  1998  &  Beiersdorfer \textit{et al.} \cite{Beiersdorfer:1998:1944} \rule[-0.6em]{0pt}{0.6em}       \\

\end{longtable}

$^\dagger$ Semiempirical prediction.

\vspace*{2mm}

}
\twocolumngrid

As for the comparison with experimental results, based on Tables~\ref{tab:excitation} and \ref{tab:3P2-3P1}, we may conclude that our theoretical predictions are in good agreement with the most of them, including the ultraprecise measurements of the $2s2p \, ^3P_2 \rightarrow 2s2p \, ^3P_1$ transition in Be-like argon~\cite{Draganic:2003:183001, SoriaOrts:2006:103002}. A discrepancy with the most precise experimental value for the $2s2s \, ^1S_0 \rightarrow 2s2p \, ^3P_{1}$ excitation energy in Be-like xenon~\cite{Trabert:2003:042501}, previously noted in Ref.~\cite{Malyshev:2021:183001}, still remains. We emphasize, however, that in this case our theoretical prediction agrees with the more recent experiment~\cite{Bernhardt:2015:144008}. We also note that the measurement of the $2s2p \, ^3P_2 \rightarrow 2s2p \, ^3P_1$ transition energy in Be-like uranium deviates from our result by almost 4 times the experimental uncertainty, while, according to the analysis in Table~\ref{tab:error}, the theoretical accuracy in this case turns out to be almost 5 times higher than the experimental one. The reason for this discrepancy is unclear to us. 

Finally, in Table~\ref{tab:QED} we show a separation of the obtained theoretical predictions for the excitation and transition energies into the non-QED and QED parts. The former is obtained by diagonalizing the matrix of the effective Hamiltonian~$H_{\rm eff}$, which along with the results of the CI calculations includes only the frequency-dependent correction of the one-photon-exchange contribution (the first diagram in Fig.~\ref{fig:diagr}), the non-QED part of the nuclear recoil effect, and the nuclear polarization correction. The QED part covers the remainder, and it is obtained as the difference of the total and non-QED terms. The separation is given based on the calculations for the LDF potential. We hope that the data presented in Tables~\ref{tab:excitation}, \ref{tab:3P2-3P1}, and \ref{tab:QED} as well as the discrepancies with the available experimental data mentioned above will serve as a trigger for new experiments with Be-like HCI, especially in view of the prospects of these systems for rigorous tests of bound-state-QED methods.

\onecolumngrid
{
\renewcommand{\arraystretch}{1.25}

\begin{longtable}{
                  l
                  S[table-format=-2.5(2), group-separator=, table-space-text-pre=$<\,$]                
                  S[table-format=-2.5(2), group-separator=, table-space-text-pre=$<\,$] 
                  S[table-format=4.5(2), group-separator=, table-space-text-pre=$<\,$] 
                  S[table-format=4.5(2), group-separator=, table-space-text-pre=$<\,$] 
                  S[table-format=-1.5(2), group-separator=, table-space-text-pre=$<\,$] 
                  S[table-format=4.5(2), group-separator=, table-space-text-pre=$<\,$] 
                 }

\caption{\label{tab:QED} 
         Non-QED and QED contributions to the excitation and transition energies in Be-like highly-charged ions (in eV).
         See the text for details.
         }\\
               
\toprule
\toprule

   \multirow{2}{*}{Contribution}    &
   \multicolumn{1}{c}{\parbox{1.8cm}{\centering \rule{0pt}{1.2em} $2s2p\,^3P_0$}}                             &
   \multicolumn{1}{c}{\parbox{1.8cm}{\centering \rule{0pt}{1.2em} $2s2p\,^3P_1$}}                              &
   \multicolumn{1}{c}{\parbox{1.8cm}{\centering \rule{0pt}{1.2em} $2s2p\,^3P_2$}}                              &
   \multicolumn{1}{c}{\parbox{1.8cm}{\centering \rule{0pt}{1.2em} $2s2p\,^1P_1$}}                              &
   \multicolumn{1}{c}{\parbox{1.8cm}{\centering \rule{0pt}{1.2em} $2s2p\,^3P_1$}}                              &
   \multicolumn{1}{c}{\parbox{1.8cm}{\centering \rule{0pt}{1.2em} $2s2p\,^3P_2$}}                               \\
                                                   &
   \multicolumn{1}{c}{\parbox{1.8cm}{\centering \rule[-0.4em]{0pt}{0.4em} $- 2s2s\,^1S_0$}}                                                         &
   \multicolumn{1}{c}{\parbox{1.8cm}{\centering \rule[-0.4em]{0pt}{0.4em} $- 2s2s\,^1S_0$}}                                                          &
   \multicolumn{1}{c}{\parbox{1.8cm}{\centering \rule[-0.4em]{0pt}{0.4em} $- 2s2s\,^1S_0$}}                                                          &
   \multicolumn{1}{c}{\parbox{1.8cm}{\centering \rule[-0.4em]{0pt}{0.4em} $- 2s2s\,^1S_0$}}                                                          &
   \multicolumn{1}{c}{\parbox{1.8cm}{\centering \rule[-0.4em]{0pt}{0.4em} $- 2s2p\,^3P_0$}}                                                          &
   \multicolumn{1}{c}{\parbox{1.8cm}{\centering \rule[-0.4em]{0pt}{0.4em} $- 2s2p\,^3P_1$}}                                                           \\ 
        
\midrule

\endfirsthead

\caption[]{Non-QED and QED contributions \textit{(continued)}.}\\

\toprule
\toprule

   \multirow{2}{*}{Contribution}    &
   \multicolumn{1}{c}{\parbox{1.8cm}{\centering \rule{0pt}{1.2em} $2s2p\,^3P_0$}}                             &
   \multicolumn{1}{c}{\parbox{1.8cm}{\centering \rule{0pt}{1.2em} $2s2p\,^3P_1$}}                              &
   \multicolumn{1}{c}{\parbox{1.8cm}{\centering \rule{0pt}{1.2em} $2s2p\,^3P_2$}}                              &
   \multicolumn{1}{c}{\parbox{1.8cm}{\centering \rule{0pt}{1.2em} $2s2p\,^1P_1$}}                              &
   \multicolumn{1}{c}{\parbox{1.8cm}{\centering \rule{0pt}{1.2em} $2s2p\,^3P_1$}}                              &
   \multicolumn{1}{c}{\parbox{1.8cm}{\centering \rule{0pt}{1.2em} $2s2p\,^3P_2$}}                               \\
                                                   &
   \multicolumn{1}{c}{\parbox{1.8cm}{\centering \rule[-0.4em]{0pt}{0.4em} $- 2s2s\,^1S_0$}}                   &
   \multicolumn{1}{c}{\parbox{1.8cm}{\centering \rule[-0.4em]{0pt}{0.4em} $- 2s2s\,^1S_0$}}                    &
   \multicolumn{1}{c}{\parbox{1.8cm}{\centering \rule[-0.4em]{0pt}{0.4em} $- 2s2s\,^1S_0$}}                    &
   \multicolumn{1}{c}{\parbox{1.8cm}{\centering \rule[-0.4em]{0pt}{0.4em} $- 2s2s\,^1S_0$}}                    &
   \multicolumn{1}{c}{\parbox{1.8cm}{\centering \rule[-0.4em]{0pt}{0.4em} $- 2s2p\,^3P_0$}}                    &
   \multicolumn{1}{c}{\parbox{1.8cm}{\centering \rule[-0.4em]{0pt}{0.4em} $- 2s2p\,^3P_1$}}                     \\ 
       
\midrule  

\endhead
\endfoot
\bottomrule
\bottomrule
\endlastfoot
       
  \multicolumn{7}{c}{\rule{0pt}{3.1ex} $Z=18$}   \\ 
 $E_{\text{non-QED}}$  &      28.46821  &      29.35659  &      31.43689  &      56.18347  &       0.88838  &       2.08030    \\ 
 $E_{\rm QED}$         &      -0.11418  &      -0.11232  &      -0.10736  &      -0.11548  &       0.00185  &       0.00497    \\ 
 $E_{\rm total}$ \rule[-0.6em]{0pt}{0.6em}      &  28.35403(41)  &  29.24427(59)  &  31.32954(39)  &  56.06799(70)  &   0.89024(39)  &   2.08527(43)    \\ 
  \multicolumn{7}{c}{\rule{0pt}{3.1ex} $Z=36$}   \\ 
 $E_{\text{non-QED}}$  &       64.0523  &       74.4028  &      126.9509  &      171.7563  &       10.3504  &       52.5481    \\ 
 $E_{\rm QED}$         &       -1.4217  &       -1.4166  &       -1.2937  &       -1.3363  &        0.0052  &        0.1229    \\ 
 $E_{\rm total}$ \rule[-0.6em]{0pt}{0.6em}      &   62.6306(21)  &   72.9862(20)  &  125.6572(19)  &  170.4200(21)  &   10.3556(18)  &   52.6710(19)    \\ 
  \multicolumn{7}{c}{\rule{0pt}{3.1ex} $Z=42$}   \\ 
 $E_{\text{non-QED}}$  &       77.7434  &       92.4616  &      200.2050  &      250.7716  &       14.7182  &      107.7434    \\ 
 $E_{\rm QED}$         &       -2.4558  &       -2.4567  &       -2.2186  &       -2.2726  &       -0.0008  &        0.2380    \\ 
 $E_{\rm total}$ \rule[-0.6em]{0pt}{0.6em}      &   75.2876(32)  &   90.0049(32)  &  197.9864(29)  &  248.4990(30)  &   14.7174(29)  &  107.9815(31)    \\ 
  \multicolumn{7}{c}{\rule{0pt}{3.1ex} $Z=54$}   \\ 
 $E_{\text{non-QED}}$  &      110.4804  &      133.2705  &      474.8108  &      538.2090  &       22.7901  &      341.5402    \\ 
 $E_{\rm QED}$         &       -5.9500  &       -5.9708  &       -5.3276  &       -5.4082  &       -0.0208  &        0.6433    \\ 
 $E_{\rm total}$ \rule[-0.6em]{0pt}{0.6em}      &  104.5305(74)  &  127.2997(73)  &  469.4832(70)  &  532.8009(69)  &   22.7693(63)  &  342.1835(73)    \\ 
  \multicolumn{7}{c}{\rule{0pt}{3.1ex} $Z=82$}   \\ 
 $E_{\text{non-QED}}$  &       234.426  &       271.385  &      2608.929  &      2711.899  &        36.959  &      2337.544    \\ 
 $E_{\rm QED}$         &       -26.353  &       -26.443  &       -24.136  &       -24.323  &        -0.090  &         2.307    \\ 
 $E_{\rm total}$ \rule[-0.6em]{0pt}{0.6em}      &   208.073(53)  &   244.942(52)  &  2584.793(51)  &  2687.576(51)  &    36.869(26)  &  2339.851(35)    \\ 
  \multicolumn{7}{c}{\rule{0pt}{3.1ex} $Z=83$}   \\ 
 $E_{\text{non-QED}}$  &       240.497  &       277.823  &      2754.137  &      2858.825  &        37.326  &      2476.314    \\ 
 $E_{\rm QED}$         &       -27.541  &       -27.635  &       -25.285  &       -25.478  &        -0.093  &         2.349    \\ 
 $E_{\rm total}$ \rule[-0.6em]{0pt}{0.6em}      &   212.956(59)  &   250.188(58)  &  2728.852(57)  &  2833.347(57)  &    37.232(27)  &  2478.664(36)    \\ 
  \multicolumn{7}{c}{\rule{0pt}{3.1ex} $Z=90$}   \\ 
 $E_{\text{non-QED}}$  &        285.02  &        324.52  &       3986.12  &       4103.58  &        39.493  &      3661.599    \\ 
 $E_{\rm QED}$         &        -37.02  &        -37.14  &        -34.72  &        -34.95  &        -0.114  &         2.421    \\ 
 $E_{\rm total}$ \rule[-0.6em]{0pt}{0.6em}      &    248.00(14)  &    287.38(14)  &   3951.40(13)  &   4068.64(13)  &    39.379(37)  &  3664.020(52)    \\ 
  \multicolumn{7}{c}{\rule{0pt}{3.1ex} $Z=92$}   \\ 
 $E_{\text{non-QED}}$  &        298.20  &        338.16  &       4418.55  &       4539.93  &        39.961  &      4080.390    \\ 
 $E_{\rm QED}$         &        -40.13  &        -40.25  &        -37.92  &        -38.16  &        -0.120  &         2.334    \\ 
 $E_{\rm total}$ \rule[-0.6em]{0pt}{0.6em}      &    258.07(11)  &    297.91(11)  &   4380.63(10)  &   4501.77(10)  &    39.841(41)  &  4082.725(56)    \\ 

\end{longtable}

}
\twocolumngrid


\section{Summary \label{sec:3}}

In the present work, the \textit{ab initio} QED approach has been applied to calculate the energies of the $n=2$ singly excited states in selected Be-like ions in a wide range of nuclear charge numbers: $18 \leqslant Z \leqslant 92$. The $2s2s \, ^1S_0$ ground state, with respect to which the excited-state energies are determined, is treated together with the close states of the same symmetry within the QED perturbation theory for quasidegenerate levels. This is done in order to properly describe the impact of the proximity of the levels and their mixing due to the electron-electron correlations on the QED effects. A similar approach is used to study the pair of the $2s2p \, ^3P_{1}$ and $2s2p \, ^1P_1$ states. The remaining excited states, $2s2p \, ^3P_{0}$ and $2s2p \, ^3P_{2}$, are considered utilizing the standard QED perturbation theory for a single level. The employed approach rigorously takes into account all the first- and second-order QED contributions in the framework of the Furry picture. The third- and higher-order interelectronic-interaction effects are evaluated within the Breit approximation. The model-QED-operator approach is used to estimate the higher-order screened QED effects. In addition, the nuclear-recoil and nuclear-polarization effects are considered. The calculations are supplemented by a diligent analysis of all possible sources of uncertainties. As a result, the most accurate theoretical predictions for the low-lyin excited states have been obtained. In general, perfect agreement is found with the results of high-precision measurements. The calculations performed may serve as a benchmark for the future theoretical and experimental studies of Be-like highly-charged ions.


\section*{Acknowledgments}

The work was supported by the Russian Science Foundation (Grant No. 22-62-00004, https://rscf.ru/project/22-62-00004/).



\begin{thebibliography}{100}

\bibitem{Beyer:2003:book:eng}
H.~F.{~}Beyer and V.~P.{~}Shevelko,
\newblock {\em Introduction to the Physics of Highly Charged Ions},
\newblock Institute of Physics Publishing, Bristol and Philadelphia, 2003.

\bibitem{Sapirstein:2008:25}
J.{~}Sapirstein and K.~T.{~}Cheng, Tests of quantum electrodynamics with
  {EBIT},
\newblock Can. J. Phys. {\bf 86},~25 (2008).

\bibitem{Beiersdorfer:2010:074032}
P.{~}Beiersdorfer, Testing {QED} and atomic-nuclear interactions with
  high-{$Z$} ions,
\newblock J. Phys. B: At. Mol. Opt. Phys. {\bf 43},~074032 (2010).

\bibitem{Glazov:2011:71}
D.~A.{~}Glazov, Y.~S.{~}Kozhedub, A.~V.{~}Maiorova, V.~M.{~}Shabaev,
  I.~I.{~}Tupitsyn, A.~V.{~}Volotka, C.{~}Kozhuharov, G.{~}Plunien, and {\relax
  Th}.{~}St{\"o}hlker, Tests of fundamental theories with heavy ions at
  low-energy regime,
\newblock Hyp. Interact. {\bf 199},~71 (2011).

\bibitem{Volotka:2012:073001}
A.~V.{~}Volotka, D.~A.{~}Glazov, O.~V.{~}Andreev, V.~M.{~}Shabaev,
  I.~I.{~}Tupitsyn, and G.{~}Plunien, Test of many-electron {QED} effects in
  the hyperfine splitting of heavy high-{$Z$} ions,
\newblock Phys. Rev. Lett. {\bf 108},~073001 (2012).

\bibitem{Shabaev:2015:031205}
V.~M.{~}Shabaev, D.~A.{~}Glazov, G.{~}Plunien, and A.~V.{~}Volotka, Theory of
  bound-electron $g$ factor in highly charged ions,
\newblock J. Phys. Chem. Ref. Data {\bf 44},~031205 (2015).

\bibitem{Shabaev:2018:60}
V.~M.{~}Shabaev, A.~I.{~}Bondarev, D.~A.{~}Glazov, M.~Y.{~}Kaygorodov,
  Y.~S.{~}Kozhedub, I.~A.{~}Maltsev, A.~V.{~}Malyshev, R.~V.{~}Popov,
  I.~I.{~}Tupitsyn, and N.~A.{~}Zubova, Stringent tests of {QED} using highly
  charged ions,
\newblock Hyp. Interact. {\bf 239},~60 (2018).

\bibitem{Kozlov:2018:045005}
M.~G.{~}Kozlov, M.~S.{~}Safronova, J.~R.{~}{Crespo L{\'o}pez-Urrutia}, and
  P.~O.{~}Schmidt, Highly charged ions: Optical clocks and applications in
  fundamental physics,
\newblock Rev. Mod. Phys. {\bf 90},~045005 (2018).

\bibitem{Indelicato:2019:232001}
P.{~}Indelicato, {QED} tests with highly charged ions,
\newblock J. Phys. B: At. Mol. Opt. Phys. {\bf 52},~232001 (2019).

\bibitem{Shabaev:2024:94:inbook}
V.~M.{~}Shabaev,
\newblock Quantum electrodynamics effects in atoms and molecules,
\newblock in {\em Comprehensive Computational Chemistry (First Edition)},
  edited by M.{~}Y{\'a}{\~n}ez and R.~J.{~}Boyd, pp. 94--128, Elsevier, Oxford,
  2024.

\bibitem{Sturm:2014:467}
S.{~}Sturm, F.{~}K{\"o}hler, J.{~}Zatorski, A.{~}Wagner, Z.{~}Harman,
  G.{~}Werth, W.{~}Quint, C.~H.{~}Keitel, and K.{~}Blaum, High-precision
  measurement of the atomic mass of the electron,
\newblock Nature {\bf 506},~467 (2014).

\bibitem{Shabaev:2006:253002}
V.~M.{~}Shabaev, D.~A.{~}Glazov, N.~S.{~}Oreshkina, A.~V.{~}Volotka,
  G.{~}Plunien, H.-J.{~}Kluge, and W.{~}Quint, $g$-factor of heavy ions: A new
  access to the fine structure constant,
\newblock Phys. Rev. Lett. {\bf 96},~253002 (2006).

\bibitem{Volotka:2014:023002}
A.~V.{~}Volotka and G.{~}Plunien, Nuclear polarization study: New frontiers for
  tests of {QED} in heavy highly charged ions,
\newblock Phys. Rev. Lett. {\bf 113},~023002 (2014).

\bibitem{Yerokhin:2016:100801}
V.~A.{~}Yerokhin, E.{~}Berseneva, Z.{~}Harman, I.~I.{~}Tupitsyn, and
  C.~H.{~}Keitel, $g$ factor of light ions for an improved determination of the
  fine-structure constant,
\newblock Phys. Rev. Lett. {\bf 116},~100801 (2016).

\bibitem{Andreev:2005:243002}
O.~{\relax Yu}.{~}Andreev, L.~N.{~}Labzowsky, G.{~}Plunien, and G.{~}Soff,
  Testing the time dependence of fundamental constants in the spectra of
  multicharged ions,
\newblock Phys. Rev. Lett. {\bf 94},~243002 (2005).

\bibitem{Berengut:2010:120801}
J.~C.{~}Berengut, V.~A.{~}Dzuba, and V.~V.{~}Flambaum, Enhanced laboratory
  sensitivity to variation of the fine-structure constant using highly charged
  ions,
\newblock Phys. Rev. Lett. {\bf 105},~120801 (2010).

\bibitem{Oreshkina:2017:030501}
N.~S.{~}Oreshkina, S.~M.{~}Cavaletto, N.{~}Michel, Z.{~}Harman, and
  C.~H.{~}Keitel, Hyperfine splitting in simple ions for the search of the
  variation of fundamental constants,
\newblock Phys. Rev. A {\bf 96},~030501 (2017).

\bibitem{Shabaev:2022:043001}
V.~M.{~}Shabaev, D.~A.{~}Glazov, A.~M.{~}Ryzhkov, C.{~}Brandau, G.{~}Plunien,
  W.{~}Quint, A.~M.{~}Volchkova, and D.~V.{~}Zinenko, Ground-state $g$ factor
  of highly charged $^{229}\mathrm{Th}$ ions: An access to the {M1} transition
  probability between the isomeric and ground nuclear states,
\newblock Phys. Rev. Lett. {\bf 128},~043001 (2022).

\bibitem{King:2022:43}
S.~A.{~}King, L.~J.{~}Spie{\ss}, P.{~}Micke, A.{~}Wilzewski, T.{~}Leopold,
  E.{~}Benkler, R.{~}Lange, N.{~}Huntemann, A.{~}Surzhykov, V.~A.{~}Yerokhin,
  J.~R.{~}{Crespo L{\'o}pez-Urrutia}, and P.~O.{~}Schmidt, An optical atomic
  clock based on a highly charged ion,
\newblock Nature {\bf 611},~43 (2022).

\bibitem{Schweppe:1991:1434}
J.{~}Schweppe, A.{~}Belkacem, L.{~}Blumenfeld, N.{~}Claytor, B.{~}Feinberg,
  H.{~}Gould, V.~E.{~}Kostroun, L.{~}Levy, S.{~}Misawa, J.~R.{~}Mowat, and
  M.~H.{~}Prior, Measurement of the {L}amb shift in lithiumlike uranium
  {(${\mathrm{U}}^{89+}$)},
\newblock Phys. Rev. Lett. {\bf 66},~1434 (1991).

\bibitem{Brandau:2003:073202}
C.{~}Brandau, C.{~}Kozhuharov, A.{~}M{\"u}ller, W.{~}Shi, S.{~}Schippers,
  T.{~}Bartsch, S.{~}B{\"o}hm, C.{~}B{\"o}hme, A.{~}Hoffknecht, H.{~}Knopp,
  N.{~}Gr{\"u}n, W.{~}Scheid, T.{~}Steih, F.{~}Bosch, B.{~}Franzke,
  P.~H.{~}Mokler, F.{~}Nolden, M.{~}Steck, T.{~}St{\"o}hlker, and
  Z.{~}Stachura, Precise determination of the
  $2{s}_{1/2}$\ensuremath{-}$2{p}_{1/2}$ splitting in very heavy lithiumlike
  ions utilizing dielectronic recombination,
\newblock Phys. Rev. Lett. {\bf 91},~073202 (2003).

\bibitem{Beiersdorfer:2005:233003}
P.{~}Beiersdorfer, H.{~}Chen, D.~B.{~}Thorn, and E.{~}Tr{\"a}bert, Measurement
  of the two-loop {L}amb shift in lithiumlike {${\mathrm{U}}^{89+}$},
\newblock Phys. Rev. Lett. {\bf 95},~233003 (2005).

\bibitem{Kozhedub:2008:032501}
Y.~S.{~}Kozhedub, O.~V.{~}Andreev, V.~M.{~}Shabaev, I.~I.{~}Tupitsyn,
  C.{~}Brandau, C.{~}Kozhuharov, G.{~}Plunien, and T.{~}St{\"o}hlker, Nuclear
  deformation effect on the binding energies in heavy ions,
\newblock Phys. Rev. A {\bf 77},~032501 (2008).

\bibitem{Sapirstein:2011:012504}
J.{~}Sapirstein and K.~T.{~}Cheng, {$S$}-matrix calculations of energy levels
  of the lithium isoelectronic sequence,
\newblock Phys. Rev. A {\bf 83},~012504 (2011).

\bibitem{Malyshev:2021:183001}
A.~V.{~}Malyshev, D.~A.{~}Glazov, Y.~S.{~}Kozhedub, I.~S.{~}Anisimova,
  M.~Y.{~}Kaygorodov, V.~M.{~}Shabaev, and I.~I.{~}Tupitsyn, \textit{Ab initio}
  calculations of energy levels in {Be}-like xenon: Strong interference between
  electron-correlation and {QED} effects,
\newblock Phys. Rev. Lett. {\bf 126},~183001 (2021).

\bibitem{Malyshev:2023:042806}
A.~V.{~}Malyshev, Y.~S.{~}Kozhedub, and V.~M.{~}Shabaev, \textit{Ab initio}
  calculations of the $2{p}_{3/2}\ensuremath{\rightarrow}2s$ transition in
  \text{{He}-,} {Li}-, and {Be}-like uranium,
\newblock Phys. Rev. A {\bf 107},~042806 (2023).

\bibitem{Loetzsch:2024:673}
R.{~}Loetzsch, H.~F.{~}Beyer, L.{~}Duval, U.{~}Spillmann, D.{~}Bana{\'s},
  P.{~}Dergham, F.~M.{~}Kr{\"o}ger, J.{~}Glorius, R.~E.{~}Grisenti,
  M.{~}Guerra, A.{~}Gumberidze, R.{~}He{\ss}, P.-M.{~}Hillenbrand,
  P.{~}Indelicato, P.{~}Jagodzinski, E.{~}Lamour, B.{~}Lorentz, S.{~}Litvinov,
  Y.~A.{~}Litvinov, J.{~}Machado, N.{~}Paul, G.~G.{~}Paulus, N.{~}Petridis,
  J.~P.{~}Santos, M.{~}Scheidel, R.~S.{~}Sidhu, M.{~}Steck, S.{~}Steydli,
  K.{~}Szary, S.{~}Trotsenko, I.{~}Uschmann, G.{~}Weber, {\relax
  Th}.{~}St{\"o}hlker, and M.{~}Trassinelli, Testing quantum electrodynamics in
  extreme fields using helium-like uranium,
\newblock Nature {\bf 625},~673 (2024).

\bibitem{Armstrong:1976:1114}
L.{~}Armstrong, W.~R.{~}Fielder, and D.~L.{~}Lin, Relativistic effects on
  transition probabilities in the {Li} and {Be} isoelectronic sequences,
\newblock Phys. Rev. A {\bf 14},~1114 (1976).

\bibitem{Braun:1984:book:rus2eng}
M.~A.{~}Braun, A.~D.{~}Gurchumelia, and U.~I.{~}Safronova,
\newblock {\em Relativistic Theory of Atom},
\newblock Nauka, Moscow, 1984.

\bibitem{Cheng:1979:111}
K.~T.{~}Cheng, Y.~K.{~}Kim, and J.~P.{~}Desclaux, Electric dipole, quadrupole,
  and magnetic dipole transition probabilities of ions isoelectronic to the
  first-row atoms, {Li} through {F},
\newblock At. Data Nucl. Data Tables {\bf 24},~111 (1979).

\bibitem{Edlen:1983:51}
B.{~}Edl{\'e}n, Comparison of theoretical and experimental level values of the
  $n = 2$ complex in ions isoelectronic with {Li}, {Be}, {O} and {F},
\newblock Phys. Scr. {\bf 28},~51 (1983).

\bibitem{Lindroth:1992:2771}
E.{~}Lindroth and J.{~}Hvarfner, Relativistic calculation of the
  {$2\,^{1}{\mathit{S}}_{0}$\ensuremath{-}$2\,^{1,3}{\mathit{P}}_{1}$}
  transitions in berylliumlike molybdenum and berylliumlike iron,
\newblock Phys. Rev. A {\bf 45},~2771 (1992).

\bibitem{Marques:1993:929}
J.~P.{~}Marques, F.{~}Parente, and P.{~}Indelicato, Hyperfine quenching of the
  $1s^2 2s2p\, ^3{P}_0$ level in berylliumlike ions,
\newblock Phys. Rev. A {\bf 47},~929 (1993).

\bibitem{Zhu:1994:3818}
X.-W.{~}Zhu and K.~T.{~}Chung, Energies and fine structures of
  $1{\mathit{s}}^{2}2snp$ ($n=2,3$) $^{1}{P}^{\mathit{o}}$ and
  $^{3}{P}_{2,1,0}^{\mathit{o}}$ states of {Be}-like ions,
\newblock Phys. Rev. A {\bf 50},~3818 (1994).

\bibitem{Safronova:1996:4036}
M.~S.{~}Safronova, W.~R.{~}Johnson, and U.~I.{~}Safronova, Relativistic
  many-body calculations of the energies of $n=2$ states for the berylliumlike
  isoelectronic sequence,
\newblock Phys. Rev. A {\bf 53},~4036 (1996).

\bibitem{Chen:1997:166}
M.~H.{~}Chen and K.~T.{~}Cheng, Energy levels of the ground state and the
  {$2s2p\,(J=1)$} excited states of berylliumlike ions: A large-scale,
  relativistic configuration-interaction calculation,
\newblock Phys. Rev. A {\bf 55},~166 (1997).

\bibitem{Santos:1998:149}
J.{~}Santos, J.{~}Marques, F.{~}Parente, E.{~}Lindroth, S.{~}Boucard, and
  P.{~}Indelicato, Multiconfiguration {D}irac-{F}ock calculation of transition
  energies in highly ionized bismuth, thorium, and uranium,
\newblock Eur. Phys. J. D {\bf 1},~149 (1998).

\bibitem{Cheng:2000:054501}
K.~T.{~}Cheng, M.~H.{~}Chen, and J.{~}Sapirstein, Quantum electrodynamic
  corrections in high-{$Z$} {Li}-like and {Be}-like ions,
\newblock Phys. Rev. A {\bf 62},~054501 (2000).

\bibitem{Safronova:2000:1213}
U.~I.{~}Safronova, Excitation energies and transition rates in
  {$\mathrm{Be}$}-, {$\mathrm{Mg}$}-, and {$\mathrm{Zn}$}-like ions,
\newblock Mol. Phys. {\bf 98},~1213 (2000).

\bibitem{Majumder:2000:042508}
S.{~}Majumder and B.~P.{~}Das, Relativistic magnetic quadrupole transitions in
  {Be}-like ions,
\newblock Phys. Rev. A {\bf 62},~042508 (2000).

\bibitem{Dong:2001:294}
C.~Z.{~}Dong, S.{~}Fritzsche, B.{~}Fricke, and W.-D.{~}Sepp, Ab-initio
  calculations for forbidden {M1} transitions in {Ar$^{13+}$} and {Ar$^{14+}$}
  ions,
\newblock Phys. Scr. {\bf 2001},~294 (2001).

\bibitem{Draganic:2003:183001}
I.{~}Dragani{\'c}, J.~R.{~}{Crespo L{\'o}pez-Urrutia}, R.{~}DuBois,
  S.{~}Fritzsche, V.~M.{~}Shabaev, R.{~}Soria~Orts, I.~I.{~}Tupitsyn, Y.{~}Zou,
  and J.{~}Ullrich, High precision wavelength measurements of {QED}-sensitive
  forbidden transitions in highly charged argon ions,
\newblock Phys. Rev. Lett. {\bf 91},~183001 (2003).

\bibitem{Gu:2005:267}
M.~F.{~}Gu, Energies of $1s^2\,2l^q \, (1 \leqslant q \leqslant 8)$ states for
  {$Z \leqslant 60$} with a combined configuration interaction and many-body
  perturbation theory approach,
\newblock At. Data Nucl. Data Tables {\bf 89},~267 (2005).

\bibitem{SoriaOrts:2006:103002}
R.{~}Soria~Orts, Z.{~}Harman, J.~R.{~}{Crespo L{\'o}pez-Urrutia},
  A.~N.{~}Artemyev, H.{~}Bruhns, A.~J.~G.{~}Mart{\'i}nez, U.~D.{~}Jentschura,
  C.~H.{~}Keitel, A.{~}Lapierre, V.{~}Mironov, V.~M.{~}Shabaev, H.{~}Tawara,
  I.~I.{~}Tupitsyn, J.{~}Ullrich, and A.~V.{~}Volotka, Exploring relativistic
  many-body recoil effects in highly charged ions,
\newblock Phys. Rev. Lett. {\bf 97},~103002 (2006).

\bibitem{Ho:2006:022510}
H.~C.{~}Ho, W.~R.{~}Johnson, S.~A.{~}Blundell, and M.~S.{~}Safronova,
  Third-order many-body perturbation theory calculations for the beryllium and
  magnesium isoelectronic sequences,
\newblock Phys. Rev. A {\bf 74},~022510 (2006).

\bibitem{Cheng:2008:052504}
K.~T.{~}Cheng, M.~H.{~}Chen, and W.~R.{~}Johnson, Hyperfine quenching of the
  {$2s2p\phantom{\rule{0.3em}{0ex}}^{3}P_{0}$} state of berylliumlike ions,
\newblock Phys. Rev. A {\bf 77},~052504 (2008).

\bibitem{Winters:2011:014013}
D.~F.~A.{~}Winters, T.{~}K{\"u}hl, D.~H.{~}Schneider, P.{~}Indelicato,
  R.{~}Reuschl, R.{~}Schuch, E.{~}Lindroth, and T.{~}St{\"o}hlker, Laser
  spectroscopy of the $(1s^22s2p) ^3{P}_0–^3{P}_1$ level splitting in
  {Be}-like krypton,
\newblock Phys. Scr. {\bf 2011},~014013 (2011).

\bibitem{Sampaio:2013:014015}
J.~M.{~}Sampaio, F.{~}Parente, C.{~}Naz{\'e}, M.{~}Godefroid, P.{~}Indelicato,
  and J.~P.{~}Marques, Relativistic calculations of $1s^2 2s2p$ level splitting
  in {Be}-like {$\mathrm{Kr}$},
\newblock Phys. Scr. {\bf T156},~014015 (2013).

\bibitem{Yerokhin:2014:022509}
V.~A.{~}Yerokhin, A.{~}Surzhykov, and S.{~}Fritzsche, Relativistic
  configuration-interaction calculation of {$K\ensuremath{\alpha}$} transition
  energies in berylliumlike iron,
\newblock Phys. Rev. A {\bf 90},~022509 (2014).

\bibitem{Yerokhin:2015:054003}
V.~A.{~}Yerokhin, A.{~}Surzhykov, and S.{~}Fritzsche, Relativistic
  configuration-interaction calculation of {$K\alpha$} transition energies in
  beryllium-like argon,
\newblock Phys. Scr. {\bf 90},~054003 (2015).

\bibitem{Wang:2015:16}
K.{~}Wang, X.~L.{~}Guo, H.~T.{~}Liu, D.~F.{~}Li, F.~Y.{~}Long, X.~Y.{~}Han,
  B.{~}Duan, J.~G.{~}Li, M.{~}Huang, Y.~S.{~}Wang, {R. Hutton}, Y.~M.{~}Zou,
  J.~L.{~}Zeng, C.~Y.{~}Chen, and J.{~}Yan, Systematic calculations of energy
  levels and transition rates of {Be}-like ions with {$Z=10-30$} using a
  combined configuration interaction and many-body perturbation theory
  approach,
\newblock Astrophys. J. Suppl. Ser. {\bf 218},~16 (2015).

\bibitem{El-Maaref:2015:2}
A.~A.{~}{El-Maaref}, S.{~}Schippers, and A.{~}M{\"u}ller, Ab-initio
  calculations of level energies, oscillator strengths and radiative rates for
  {E1} transitions in beryllium-like iron,
\newblock Atoms {\bf 3},~2 (2015).

\bibitem{Li:2017:720}
K.~K.{~}Li, L.{~}Zhuo, C.~M.{~}Zhang, C.{~}Chen, and B.~C.{~}Gou, Energies and
  hyperfine structures of the {$1s^22s2p\, ^3P^o$} state of {Be}-like ions with
  {$Z=11-18$},
\newblock Can. J. Phys. {\bf 95},~720 (2017).

\bibitem{Kaygorodov:2019:032505}
M.~Y.{~}Kaygorodov, Y.~S.{~}Kozhedub, I.~I.{~}Tupitsyn, A.~V.{~}Malyshev,
  D.~A.{~}Glazov, G.{~}Plunien, and V.~M.{~}Shabaev, Relativistic calculations
  of the ground and inner-${L}$-shell excited energy levels of berylliumlike
  ions,
\newblock Phys. Rev. A {\bf 99},~032505 (2019).

\bibitem{Malyshev:2021:652}
A.~V.{~}Malyshev, Y.~S.{~}Kozhedub, I.~S.{~}Anisimova, D.~A.{~}Glazov,
  M.~Y.{~}Kaygorodov, I.~I.{~}Tupitsyn, and V.~M.{~}Shabaev, Binding energy of
  the ground state of beryllium-like molybdenum: Correlation and
  quantum-electrodynamic effects,
\newblock Opt. Spectrosc. {\bf 129},~652 (2021).

\bibitem{Widing:1975:L33}
K.~G.{~}Widing, {Fe XXIII} 263 {{\r{A}}} and {Fe XXIV} 255 {{\r{A}}} emission
  in solar flares,
\newblock Astrophys. J. {\bf 197},~L33 (1975).

\bibitem{Dere:1978:1062}
K.~P.{~}Dere, Spectral lines observed in solar flares between 171 and 630
  angstroms,
\newblock Astrophys. J. {\bf 221},~1062 (1978).

\bibitem{Denne:1989:1488}
B.{~}Denne, E.{~}Hinnov, J.{~}Ramette, and B.{~}Saoutic, Spectrum lines of {Kr
  XXVIII} $-$ {Kr XXXIV} observed in the {JET} tokamak,
\newblock Phys. Rev. A {\bf 40},~1488 (1989).

\bibitem{Denne:1989:3702}
B.{~}Denne, G.{~}Magyar, and J.{~}Jacquinot, Berylliumlike {Mo XXXIX} and
  lithiumlike {Mo XL} observed in the {J}oint {E}uropean {T}orus tokamak,
\newblock Phys. Rev. A {\bf 40},~3702 (1989).

\bibitem{Martin:1990:6570}
S.{~}Martin, A.{~}Denis, M.~C.{~}{Buchet-Poulizac}, J.~P.{~}Buchet, and
  J.{~}D{\'e}sesquelles, $2s-2p$ transitions in heliumlike and lithiumlike
  krypton,
\newblock Phys. Rev. A {\bf 42},~6570 (1990).

\bibitem{Sugar:1991:859}
J.{~}Sugar and A.{~}Musgrove, Energy levels of krypton, {Kr I} through {Kr
  XXXVI},
\newblock J. Phys. Chem. Ref. Data {\bf 20},~859 (1991).

\bibitem{Beiersdorfer:1993:3939}
P.{~}Beiersdorfer, D.{~}Knapp, R.~E.{~}Marrs, S.~R.{~}Elliott, and
  M.~H.{~}Chen, Structure and {L}amb shift of
  2${\mathit{s}}_{1/2}$\ensuremath{-}2${\mathit{p}}_{3/2}$ levels in
  lithiumlike $\mathrm{U}^{89+}$ through neonlike $\mathrm{U}^{82+}$,
\newblock Phys. Rev. Lett. {\bf 71},~3939 (1993).

\bibitem{Beiersdorfer:1995:2693}
P.{~}Beiersdorfer, A.{~}Osterheld, S.~R.{~}Elliott, M.~H.{~}Chen, D.{~}Knapp,
  and K.{~}Reed, Structure and {L}amb shift of
  2${\mathit{s}}_{1/2}$\ensuremath{-}2${\mathit{p}}_{3/2}$ levels in
  lithiumlike $\mathrm{Th}^{87+}$ through neonlike $\mathrm{Th}^{80+}$,
\newblock Phys. Rev. A {\bf 52},~2693 (1995).

\bibitem{Bieber:1997:64}
D.~J.{~}Bieber, H.~S.{~}Margolis, P.~K.{~}Oxley, and J.~D.{~}Silver, Studies of
  magnetic dipole transitions in highly charged argon and barium using an
  electron beam ion trap,
\newblock Phys. Scr. {\bf T73},~64 (1997).

\bibitem{Beiersdorfer:1998:1944}
P.{~}Beiersdorfer, A.~L.{~}Osterheld, and S.~R.{~}Elliott, Measurements and
  modeling of electric-dipole-forbidden ${2p}_{1/2}$\ensuremath{-}$2p_{3/2}$
  transitions in fluorinelike {${\mathrm{U}}^{81+}$} through berylliumlike
  {${\mathrm{U}}^{88+}$},
\newblock Phys. Rev. A {\bf 58},~1944 (1998).

\bibitem{Trabert:2003:042501}
E.{~}Tr{\"a}bert, P.{~}Beiersdorfer, J.~K.{~}Lepson, and H.{~}Chen, Extreme
  ultraviolet spectra of highly charged {Xe} ions,
\newblock Phys. Rev. A {\bf 68},~042501 (2003).

\bibitem{Feili:2005:48}
D.{~}Feili, B.{~}Zimmermann, C.{~}Neacsu, {\relax Ph}.{~}Bosselmann,
  K.-H.{~}Schartner, F.{~}Folkmann, A.~E.{~}Livingston, E.{~}Tr{\"a}bert, and
  P.~H.{~}Mokler, {$2s^2 \,^1S_0$\ensuremath{-}$2s2p \,^3P_1$} intercombination
  transition wavelengths in {Be}-like {$\mathrm{Ag}^{43+}$},
  {$\mathrm{Sn}^{46+}$}, and {$\mathrm{Xe}^{50+}$} ions,
\newblock Phys. Scr. {\bf 71},~48 (2005).

\bibitem{Katai:2007:120}
R.{~}Katai, S.{~}Morita, and M.{~}Goto, Identification and intensity analysis
  on forbidden magnetic dipole emission lines of highly charged {Al}, {Ar},
  {Ti} and {Fe} ions in {LHD},
\newblock J. Quant. Spectrosc. Radiat. Transf. {\bf 107},~120 (2007).

\bibitem{Saloman:2010:033101}
E.~B.{~}Saloman, Energy levels and observed spectral lines of ionized argon,
  {ArII} through {ArXVIII},
\newblock J. Phys. Chem. Ref. Data {\bf 39},~033101 (2010).

\bibitem{Bernhardt:2015:144008}
D.{~}Bernhardt, C.{~}Brandau, Z.{~}Harman, C.{~}Kozhuharov, S.{~}B{\"o}hm,
  F.{~}Bosch, S.{~}Fritzsche, J.{~}Jacobi, S.{~}Kieslich, H.{~}Knopp,
  F.{~}Nolden, W.{~}Shi, Z.{~}Stachura, M.{~}Steck, {\relax
  Th}.{~}St{\"o}hlker, S.{~}Schippers, and A.{~}M{\"u}ller, Spectroscopy of
  berylliumlike xenon ions using dielectronic recombination,
\newblock J. Phys. B: At. Mol. Opt. Phys. {\bf 48},~144008 (2015).

\bibitem{Furry:1951:115}
W.~H.{~}Furry, On bound states and scattering in positron theory,
\newblock Phys. Rev. {\bf 81},~115 (1951).

\bibitem{TTGF}
V.~M.{~}Shabaev, Two-time {G}reen's function method in quantum electrodynamics
  of high-{$Z$} few-electron atoms,
\newblock Phys. Rep. {\bf 356},~119 (2002).

\bibitem{Shabaev:2005:062105}
V.~M.{~}Shabaev, I.~I.{~}Tupitsyn, K.{~}Pachucki, G.{~}Plunien, and
  V.~A.{~}Yerokhin, Radiative and correlation effects on the
  parity-nonconserving transition amplitude in heavy alkali-metal atoms,
\newblock Phys. Rev. A {\bf 72},~062105 (2005).

\bibitem{Yerokhin:2015:033103}
V.~A.{~}Yerokhin and V.~M.{~}Shabaev, {L}amb shift of $n = 1$ and $n = 2$
  states of hydrogen-like atoms, {$1 \leqslant Z \leqslant 110$},
\newblock J. Phys. Chem. Ref. Data {\bf 44},~033103 (2015).

\bibitem{Yerokhin:2018:052509}
V.~A.{~}Yerokhin, Two-loop self-energy in the {L}amb shift of the ground and
  excited states of hydrogenlike ions,
\newblock Phys. Rev. A {\bf 97},~052509 (2018).

\bibitem{Malyshev:2014:062517}
A.~V.{~}Malyshev, A.~V.{~}Volotka, D.~A.{~}Glazov, I.~I.{~}Tupitsyn,
  V.~M.{~}Shabaev, and G.{~}Plunien, {QED} calculation of the ground-state
  energy of berylliumlike ions,
\newblock Phys. Rev. A {\bf 90},~062517 (2014).

\bibitem{Malyshev:2017:022512}
A.~V.{~}Malyshev, D.~A.{~}Glazov, A.~V.{~}Volotka, I.~I.{~}Tupitsyn,
  V.~M.{~}Shabaev, G.{~}Plunien, and {\relax Th}.{~}St{\"o}hlker, Ground-state
  ionization energies of boronlike ions,
\newblock Phys. Rev. A {\bf 96},~022512 (2017).

\bibitem{Kozhedub:2019:062506}
Y.~S.{~}Kozhedub, A.~V.{~}Malyshev, D.~A.{~}Glazov, V.~M.{~}Shabaev, and
  I.~I.{~}Tupitsyn, {QED} calculation of electron-electron correlation effects
  in heliumlike ions,
\newblock Phys. Rev. A {\bf 100},~062506 (2019).

\bibitem{Malyshev:2019:010501_R}
A.~V.{~}Malyshev, Y.~S.{~}Kozhedub, D.~A.{~}Glazov, I.~I.{~}Tupitsyn, and
  V.~M.{~}Shabaev, {QED} calculations of the $n=2$ to $n=1$ x-ray transition
  energies in middle-${Z}$ heliumlike ions,
\newblock Phys. Rev. A {\bf 99},~010501(R) (2019).

\bibitem{Bratzev:1977:2655}
V.~F.{~}Bratzev, G.~B.{~}Deyneka, and I.~I.{~}Tupitsyn, Application of the
  {H}artree-{F}ock method to calculation of relativistic atomic wave functions,
\newblock Izv. Acad. Nauk SSSR, Ser. Fiz. {\bf 41},~2655 (1977),
\newblock [Bull. Acad. Sci. USSR, Phys. Ser. \textbf{41}, 173 (1977)].

\bibitem{Tupitsyn:2003:022511}
I.~I.{~}Tupitsyn, V.~M.{~}Shabaev, J.~R.{~}{Crespo L{\'o}pez-Urrutia},
  I.{~}Dragani{\'c}, R.{~}Soria~Orts, and J.{~}Ullrich, Relativistic
  calculations of isotope shifts in highly charged ions,
\newblock Phys. Rev. A {\bf 68},~022511 (2003).

\bibitem{Kozhedub:2010:042513}
Y.~S.{~}Kozhedub, A.~V.{~}Volotka, A.~N.{~}Artemyev, D.~A.{~}Glazov,
  G.{~}Plunien, V.~M.{~}Shabaev, I.~I.{~}Tupitsyn, and {\relax
  Th}.{~}St{\"o}hlker, Relativistic recoil, electron-correlation, and {QED}
  effects on the $2{p}_{j}$\ensuremath{-}$2s$ transition energies in {Li}-like
  ions,
\newblock Phys. Rev. A {\bf 81},~042513 (2010).

\bibitem{Shabaev:1985:394}
V.~M.{~}Shabaev, Mass corrections in a strong nuclear field,
\newblock Teor. Mat. Fiz. {\bf 63},~394 (1985),
\newblock [Theor. Math. Phys. \textbf{63}, 588 (1985)].

\bibitem{Shabaev:1988:107}
V.~M.{~}Shabaev, Nuclear recoil effect in relativistic theory of multiply
  charged ions,
\newblock Yad. Fiz. {\bf 47},~107 (1988),
\newblock [Sov. J. Nucl. Phys. \textbf{47}, 69 (1988)].

\bibitem{Shabaev:1998:59}
V.~M.{~}Shabaev, {QED} theory of the nuclear recoil effect in atoms,
\newblock Phys. Rev. A {\bf 57},~59 (1998).

\bibitem{Pachucki:1995:1854}
K.{~}Pachucki and H.{~}Grotch, Pure recoil corrections to hydrogen energy
  levels,
\newblock Phys. Rev. A {\bf 51},~1854 (1995).

\bibitem{Adkins:2007:042508}
G.~S.{~}Adkins, S.{~}Morrison, and J.{~}Sapirstein, Recoil corrections in
  highly charged ions,
\newblock Phys. Rev. A {\bf 76},~042508 (2007).

\bibitem{Palmer:1987:5987}
C.~W.~P.{~}Palmer, Reformulation of the theory of the mass shift,
\newblock J. Phys. B: At. Mol. Phys. {\bf 20},~5987 (1987).

\bibitem{Pachucki:2024:032804}
K.{~}Pachucki and V.~A.{~}Yerokhin, Heavy-particle quantum electrodynamics,
\newblock Phys. Rev. A {\bf 110},~032804 (2024).

\bibitem{Plunien:1995:1119:1996:4614:join_pr}
G.{~}Plunien and G.{~}Soff, Nuclear-polarization contribution to the {L}amb
  shift in actinide nuclei,
\newblock Phys. Rev. A \textbf{51}, 1119 (1995); \textbf{53}, 4614(E) (1996).

\bibitem{Nefiodov:1996:227}
A.~V.{~}Nefiodov, L.~N.{~}Labzowsky, G.{~}Plunien, and G.{~}Soff, Nuclear
  polarization effects in spectra of multicharged ions,
\newblock Phys. Lett. A {\bf 222},~227 (1996).

\bibitem{Shabaev:2013:012513}
V.~M.{~}Shabaev, I.~I.{~}Tupitsyn, and V.~A.{~}Yerokhin, Model operator
  approach to the {L}amb shift calculations in relativistic many-electron
  atoms,
\newblock Phys. Rev. A {\bf 88},~012513 (2013).

\bibitem{Shabaev:2015:175:2018:69:join_pr}
V.~M.{~}Shabaev, I.~I.{~}Tupitsyn, and V.~A.{~}Yerokhin, {QEDMOD}: {F}ortran
  program for calculating the model {L}amb-shift operator,
\newblock Comput. Phys. Commun. \textbf{189}, 175 (2015); \textbf{223}, 69
  (2018).

\bibitem{Tiesinga:2021:025010}
E.{~}Tiesinga, P.~J.{~}Mohr, D.~B.{~}Newell, and B.~N.{~}Taylor, {CODATA}
  recommended values of the fundamental physical constants: 2018,
\newblock Rev. Mod. Phys. {\bf 93},~025010 (2021).

\bibitem{Sapirstein:1999:259}
J.{~}Sapirstein, K.~T.{~}Cheng, and M.~H.{~}Chen, Potential independence of the
  solution to the relativistic many-body problem and the role of
  negative-energy states in heliumlike ions,
\newblock Phys. Rev. A {\bf 59},~259 (1999).

\bibitem{Shabaev:1993:4703}
V.~M.{~}Shabaev, {S}chrodinger-like equation for the relativistic few-electron
  atom,
\newblock J. Phys. B: At. Mol. Opt. Phys. {\bf 26},~4703 (1993).

\bibitem{Drake:1988:586}
G.~W.{~}Drake, Theoretical energies for the $n=1$ and 2 states of the helium
  isoelectronic sequence up to {$Z=100$},
\newblock Can. J. Phys. {\bf 66},~586 (1988).

\bibitem{Artemyev:2005:062104}
A.~N.{~}Artemyev, V.~M.{~}Shabaev, V.~A.{~}Yerokhin, G.{~}Plunien, and
  G.{~}Soff, {QED} calculation of the $n=1$ and $n=2$ energy levels in
  {He}-like ions,
\newblock Phys. Rev. A {\bf 71},~062104 (2005).

\bibitem{Shabaev:1998:4235}
V.~M.{~}Shabaev, A.~N.{~}Artemyev, T.{~}Beier, G.{~}Plunien, V.~A.{~}Yerokhin,
  and G.{~}Soff, Recoil correction to the ground-state energy of hydrogenlike
  atoms,
\newblock Phys. Rev. A {\bf 57},~4235 (1998).

\bibitem{Aleksandrov:2015:144004}
I.~A.{~}Aleksandrov, A.~A.{~}Shchepetnov, D.~A.{~}Glazov, and V.~M.{~}Shabaev,
  Finite nuclear size corrections to the recoil effect in hydrogenlike ions,
\newblock J. Phys. B: At. Mol. Opt. Phys. {\bf 48},~144004 (2015).

\bibitem{Anisimova:2022:062823}
I.~S.{~}Anisimova, A.~V.{~}Malyshev, D.~A.{~}Glazov, M.~Y.{~}Kaygorodov,
  Y.~S.{~}Kozhedub, G.{~}Plunien, and V.~M.{~}Shabaev, Model-{QED}-operator
  approach to relativistic calculations of the nuclear recoil effect in
  many-electron atoms and ions,
\newblock Phys. Rev. A {\bf 106},~062823 (2022).

\bibitem{Pachucki:2023:053002}
K.{~}Pachucki and V.~A.{~}Yerokhin, {QED} theory of the nuclear recoil with
  finite size,
\newblock Phys. Rev. Lett. {\bf 130},~053002 (2023).

\bibitem{Peck:1972:958}
E.~R.{~}Peck and K.{~}Reeder, Dispersion of air*,
\newblock J. Opt. Soc. Am. {\bf 62},~958 (1972).

\end{thebibliography}


\end{document}